\documentclass{aa}
\usepackage{txfonts}
\usepackage{natbib}
\usepackage{amssymb}
\usepackage{amsmath}
\bibpunct{(}{)}{;}{a}{}{,}
\usepackage{epsf}
\usepackage{graphicx}
\usepackage{epsfig}
\usepackage{graphics}
\usepackage{subfigure}
\usepackage[english]{babel}
\usepackage{rotating}
\usepackage{color}

\newcommand{\mum}{$\mu$m}
\newcommand{\teff}{$T_{\rm{eff}}$}

\newcommand{\lL}{\ifmmode \log \frac{L}{L_{\sun}} \else $\log \frac{L}{L_{\sun}}$\fi}

\newcommand{\myr}{M$_{\sun}$ yr$^{-1}$}

\newcommand{\vinf}{$v_{\infty}$}

\newcommand{\kms}{km~s$^{-1}$}
\newcommand{\msun}{M$_{\sun}$}
\newcommand{\brg}{Br$\gamma$}

\begin{document}

\title{Massive stars in the young cluster VVV~CL074\thanks{Based on observations conducted at ESO on the Very Large Telescope under program 099.D-0161(B).}}
\author{F. Martins\inst{1}
\and A.-N. Chen\'e\inst{2}
\and J.-C. Bouret\inst{3}
\and J. Borissova\inst{4,5}
\and J. Groh\inst{6}
\and S. Ram\'irez Alegr\'ia\inst{7}
\and D. Minniti\inst{8}
}
\institute{LUPM, Universit\'e de Montpellier, CNRS, Place Eug\`ene Bataillon, F-34095 Montpellier, France  \\
%           \email{fabrice.martins@umontpellier.fr}
\and
Gemini Observatory, Northern Operations Center, 670 N. A'ohoku Place, Hilo, Hawaii, 96720, USA\\
\and
Aix Marseille Universit\'e, CNRS, CNES, LAM, Marseille, France\\
\and
Instituto de F\'isica y Astronom\'ia, U. de Valpara\'iso, Av. Gran Breta$\tilde{n}$a 1111, Playa Ancha, 5030 Casilla, Chile\\
\and
Millennium Institute of Astrophysics, Av. Vicu\~na Mackenna 4860, 782-0436 Macul, Santiago, Chile
\and
Trinity College Dublin, The University of Dublin, Dublin 2, Ireland\\
\and
Centro de Astronomía (CITEVA), Universidad de Antofagasta, Avenida Angamos 601, Antofagasta 1270300, Chile\\
\and
Departamento de Ciencias Físicas, Facultad de Ciencias Exactas, Universidad Andrés Bello, Av. Fernández Concha 700, Las Condes, Santiago, Chile
}

\offprints{Fabrice Martins\\ \email{fabrice.martins@umontpellier.fr}}

\date{Received / Accepted }

\abstract
{The evolution of massive stars is not fully constrained. Studies of young massive clusters hosting various populations of massive stars can help refine our understanding of the life and fate of massive stars. }
{In this context, our goal is to study the massive stellar content of the young massive cluster VVV~CL074.}
{We obtained K-band spectroscopy of the brightest cluster members in order to identify the massive star population. We also determined the stellar properties of the cluster's massive stars to better quantify the evolutionary sequences linking different types of massive stars. We collected integral field spectroscopy of selected fields in the cluster VVV~CL074 with SINFONI on the ESO/VLT. We performed a spectral classification based on the K-band spectra and comparison to infrared spectral atlases. We determined the stellar parameters of the massive stars from analysis with atmosphere models computed with the code CMFGEN.}
{We uncover a population of 25 early-type (OB and Wolf-Rayet) stars, 19 being newly discovered by our observations out of which 15 are likely cluster members. The cluster's spectrophotometric distance is 10.2$\pm$1.6 kpc, placing it close to the intersection of the galactic bar and the Norma arm, beyond the galactic center. This makes VVV~CL074 one the farthest young massive clusters identified so far. Among the massive stars population, three objects are Wolf-Rayet stars, the remaining are O and B stars. From the Hertzsprung-Russell diagram we find that most stars have an age between 3 and 6 Myr according to the Geneva evolutionary tracks. WN8 and WC8-9 stars are the descendants of stars with initial masses between 40 and 60 \msun. The massive star population of VVV~CL074 is very similar to that of the cluster DBS2003-179 and to a lesser extent to that of the Quintuplet cluster, indicating the same age. The central cluster of the Galaxy is $\sim$3 Myr older. From the comparison of the massive stars populations in these four clusters, one concludes that galactic stars with an initial mass in the range 40 to 60 \msun\ likely go through a WN8-9 phase.}
{}

\keywords{Stars: massive -- Stars: early-type -- Stars: Wolf-Rayet -- Stars: atmospheres -- Stars: fundamental parameters -- Stars: evolution}

\authorrunning{Martins et al.}
\titlerunning{Massive stars in VVV~CL074}

\maketitle

%%%%%%%%%%%%%%%%%%%%%%%%%%%%%%%%%%%%%%%%%%%%%%%%%%%%%%%%%%%%%%%%%%%%%%%%%%%%%%%%%%%%%%%%%%%%%%%%%%%%%%%%%%%%%%%%%%%%%%%%%%%%%%%
%%%%%%%%%%%%%%%%%%%%%%%%%%%%%%%%%%%%%%%%%%%%%%%%%%%%%%%%%%%%%%%%%%%%%%%%%%%%%%%%%%%%%%%%%%%%%%%%%%%%%%%%%%%%%%%%%%%%%%%%%%%%%%%
\section{Introduction}
\label{s_intro}

Massive stars are born as O and B stars and live as such while they stay on the main sequence. After core-hydrogen depletion, they evolve into different types of objects depending on their initial mass. It is thought that for initial masses below $\sim$25 \msun\ massive stars become red supergiants \citep[RSG,][]{ek12,lc18} with hydrogen-rich envelopes. Subsequent evolution is relatively uncertain. The majority of type II supernovae (SN) progenitors are expected to be RSGs \citep{eldridge13,groh13,smartt15}. 
However, mass loss during the RSG phase exposes the internal, hydrogen-poor layers of the star to the surface, leading to its transformation into a Wolf-Rayet star (WR), most likely of the WN sequence (i.e., showing strong nitrogen emission lines; see, e.g., \citet{paul07}). Above 25 \msun\ the RSG phase is probably suppressed, mass loss preventing the star from evolving towards the red part of the Hertzsprung-Russell (HR) diagram. A luminous blue variable (LBV) eruptive phase may be encountered before the star turns into a WR star, again most likely of the WN sequence. WC stars, whose spectra are dominated by carbon and helium emission lines, likely correspond to subsequent phases of evolution since they show the products of helium burning (and no hydrogen) at their surface  (see, e.g., \citet{langer12}). Finally, WO stars may be the final state of evolution of the most massive stars \citep{groh14,tramper15,sander18}.

We see that while general trends in the relation between various types of massive stars can be defined, none is accurately constrained. In particular, the relations between subtypes of WR stars are poorly known. The main reason is that we do not have access to time sequences, but rather to samples of stars of different ages, and thus different initial masses. This makes it difficult to reconstruct the evolution of each type of object. One way of bypassing this difficulty is to rely on evolutionary models and to compute their spectroscopic appearance, as pioneered by \citet{sdk96} and recently revisited by \citet{groh14} and \citet{mp17}. This way a direct relation between spectral type and evolution can be identified. The drawback is that such a process heavily relies on the assumptions of both evolutionary and atmosphere models. Another approach is to study several samples of massive stars, each with a given age. This way we get snapshots of stellar evolution at different times and, post-main sequence evolution being much faster than hydrogen-burning, the progenitors and evolution of evolved massive stars can be better identified. This approach can be conducted by studying young massive clusters \citep{martins07,martins08,davies07,davies08,neg10,liermann12,davies12a,davies12b,borissova12,ram12,ram14}.  

The ESO VISTA Variables in the V\'ia L\'actea (VVV) survey is a near-infrared photometric survey of the Galaxy \citep{minniti10,saito12,hempel14}. Several young massive clusters have been discovered \citep{borissova11,chene12} and follow-up spectroscopy has allowed the identification of massive stars in a number of them \citep{chene13,chene15,ram14,ram16}. \citet{herve16} presented preliminary results on the analysis of WR stars in four clusters. In the present paper, we describe new spectroscopic observations of the young massive cluster VVV~CL074. Using integral-field spectroscopy we characterize the spectral properties of the brightest members and identify new OB and WR stars. We determine the stellar parameters of most massive stars and discuss the results in the context of defining evolutionary sequences for massive stars. We also compare VVV~CL074 with other prototypical young massive galactic clusters.

%%%%%%%%%%%%%%%%%%%%%%%%%%%%%%%%%%%%%%%%%%%%%%%%%%%%%%%%%%%%%%%%%%%%%%%%%%%%%%%%%%%%%%%%%%%%%%%%%%%%%%%%%%%%%%%%%%%%%%%%%%%%%%%
\section{Observations and data reduction}
\label{s_obs}

\begin{figure}[t]
\centering
\includegraphics[width=0.49\textwidth]{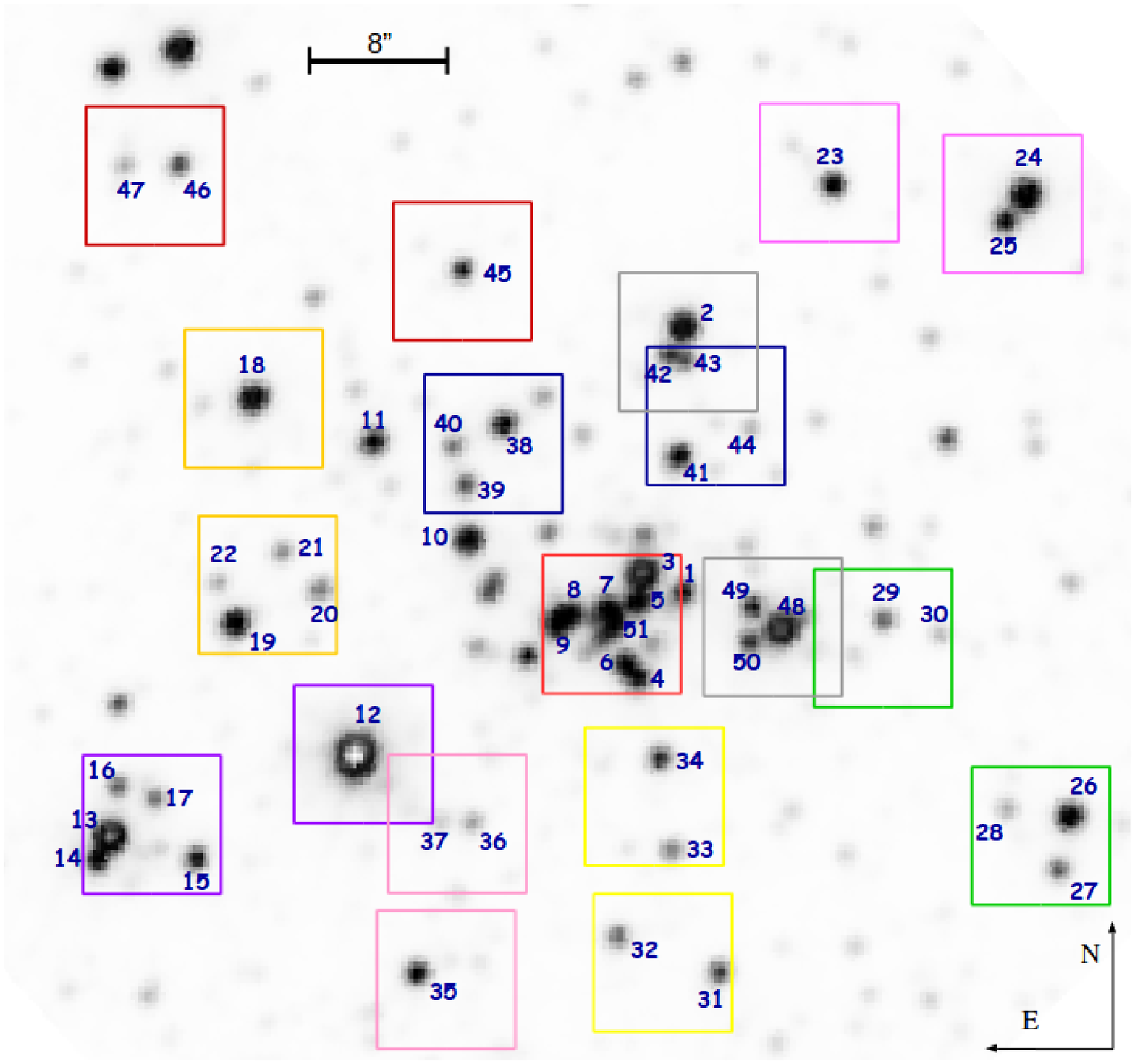}
\caption{K-band image of VVV~CL074 from the VVV survey. Observed fields (squares) and stars identificators are overplotted. Squares of the same color correspond to the same observing block. The size of the observed fields is 8\arcsec, corresponding to a linear distance of about 0.4~pc at the estimated cluster distance (10.2 kpc).}
\label{fig_chart}
\end{figure}

Observations were conducted in service mode between April 17$^{}$ and June 9$^{}$ 2017 at ESO/VLT on the integral-field spectrograph SINFONI. Our program was divided into ten observing blocks covering various cluster fields. Figure\ \ref{fig_chart} shows a sketch of these regions. Each observing block lasted one hour and was observed as a filler program since our requests on seeing were moderate. In practice, visible seeing as measured on the guide probe varied between 1.0\arcsec\ and 1.5\arcsec\ depending on the field and date of observation. The K-band setting was selected, ensuring a spectral resolution of $\sim$5000.

\begin{table*}
\begin{center}
  \caption{Spectral types, VVV photometry, and coordinates of the observed stars.} \label{tab_obs}
\begin{tabular}{lccccccc}
\hline
Star& ST              & J      &  H    & K       &  RA            &    DEC    \\    
    &                 & [mag]  & [mag] & [mag]   & [hr:m:s]       &  [d:m:s]     \\
\hline
2   &  WC8            & 16.92  & 13.23 & 10.31   &  16:32:05.28   &  -47:49:15.78  \\
3   &  WN8            & 14.72  & 11.82 & 10.17   &  16:32:05.46   &  -47:49:29.53 \\
4   &  O7-9.5 I       & 16.02  & 13.30 & 11.93   &  16:32:05.48   &  -47:49:35.41  \\
%5   &  O7-8 V        & 15.64  & 12.90 & 11.54  \\
6   &  O6-7 I         & 16.00  & 13.24 & 11.89   &  16:32:05.54   &  -47:49:34.78   \\
7   &  O6-7 III       & 15.98  & 13.46 & 12.19   &  16:32:05.67   &  -47:49:31.53  \\
8   &  CO             & 13.98  & 12.27 & 11.10   &  16:32:05.88   &  -47:49:32.16  \\
9   &  O4-6 If+       & 15.22  & 12.53 & 11.31   &  16:32:05.96   &  -47:49:32.53  \\
12  &  B0.5-2 I$^1$   & 12.34  & 10.09 &  8.89   &  16:32:07.08   &  -47:49:39.83  \\
13  &  WC9            & 15.08  & 12.13 & 10.30   &  16:32:08.43   &  -47:49:44.70  \\
14  &  CO             & 16.77  & 13.54 & 12.09   &  16:32:08.50   &  -47:49:46.08  \\
15  &  CO             & 14.31  & 12.87 & 12.29   &  16:32:07.97   &  -47:49:45.03  \\
16  &  CO             & 17.53  & 14.30 & 12.87   &  16:32:08.41   &  -47:49:41.95  \\
17  &  O7-8 V-III$^2$ & 13.79  & 13.22 & 12.99   &  16:32:08.21   &  -47:49:42.45  \\
18  &  O9.5-B2 I      & 14.46  & 12.06 & 10.84   &  16:32:07.61   &  -47:49:21.17  \\
19  &  CO             & 15.00  & 12.40 & 11.27   &  16:32:07.67   &  -47:49:33.79  \\
20  &  $>$B3          & 16.86  & 14.50 & 13.30   &  16:32:07.22   &  -47:49:31.79  \\
21  &  $>$B3$^2$      & 13.82  & 13.30 & 13.19   &  16:32:07.43   &  -47:49:29.79  \\ 
22  &  CO             & 17.84  & 14.92 & 13.67   &  16:32:07.77   &  -47:49:31.54  \\
23  &  CO             & 16.19  & 13.28 & 11.96   &  16:32:04.44   &  -47:49:07.51  \\
24  &  $>$B7 I        & 15.54  & 12.23 & 10.54   &  16:32:03.38   &  -47:49:08.01  \\
25  &  CO             & 16.70  & 13.42 & 11.95   &  16:32:03.48   &  -47:49:09.51  \\
26  &  CO             & 12.77  & 11.73 & 11.35   &  16:32:03.05   &  -47:49:42.59  \\
27  &  CO             & 18.38  & 14.49 & 12.69   &  16:32:03.12   &  -47:49:45.59  \\
28  &  B$^2$          & 14.73  & 13.92 & 13.60   &  16:32:03.40   &  -47:49:42.34  \\
29  &  CO             & 13.74  & 13.01 & 12.84   &  16:32:04.10   &  -47:49:32.21  \\
30  &  CO             & 18.64  & 15.30 & 13.73   &  16:32:03.81   &  -47:49:32.46  \\
31  &  CO             & 15.22  & 13.35 & 12.54   &  16:32:04.99   &  -47:49:51.64  \\
32  &  B              & 16.23  & 13.95 & 12.85   &  16:32:05.55   &  -47:49:49.89  \\
33  &  CO             & 16.68  & 14.14 & 13.02   &  16:32:05.27   &  -47:49:45.01  \\
34  &  early/emission & 16.45  & 13.85 & 12.63   &  16:32:05.34   &  -47:49:40.01  \\
35  &  CO             & 15.90  & 13.26 & 12.09   &  16:32:06.66   &  -47:49:52.21  \\
36  &  CO             & 16.37  & 14.23 & 13.31   &  16:32:06.37   &  -47:49:43.71  \\
37  &  other          & 17.44  & 15.22 & 14.17   &  16:32:06.55   &  -47:49:43.83  \\
38  &  CO             & 16.79  & 13.29 & 11.71   &  16:32:06.26   &  -47:49:21.33  \\
39  &  B?             & 17.11  & 14.20 & 12.79   &  16:32:06.46   &  -47:49:24.83  \\
40  &  CO             & 18.16  & 14.58 & 12.97   &  16:32:06.53   &  -47:49:22.58  \\
41  &  $>$B3$^2$      & 12.40  & 12.10 & 11.96   &  16:32:05.27   &  -47:49:22.96  \\
42  &  O7-8 III-I     & 16.89  & 14.05 & 12.64   &  16:32:05.34   &  -47:49:17.21  \\
43  &  CO             & 18.41  & 14.54 & 12.75   &  16:32:05.27   &  -47:49:17.46  \\
44  &  B              & 18.48  & 15.39 & 13.74   &  16:32:04.88   &  -47:49:21.21  \\
45  &  CO             & 16.60  & 13.71 & 12.40   &  16:32:06.52   &  -47:49:12.93  \\
46  &  CO             & 16.43  & 13.81 & 12.64   &  16:32:08.12   &  -47:49:07.18  \\
47  &  OB             & 19.05  & 15.46 & 13.64   &  16:32:08.40   &  -47:49:07.31  \\
48  &  early/emission & 14.31  & 11.67 & 10.33   &  16:32:08.68   &  -47:49:32.41  \\
49  &  O7-9.5 I$^3$   & 16.21  & 13.58 & 12.27   &  16:32:04.84   &  -47:49:31.41  \\
50  &  O7-9.5 I$^3$   & 16.53  & 13.82 & 12.49   &  16:32:04.86   &  -47:49:33.41  \\
51  &  O6-8 III       & 16.17  & 13.54 & 12.22   &  16:32:05.62   &  -47:49:32.56  \\
\hline                                                                     
\end{tabular}                                                              
\tablefoot{1: 2MASS photometry; 2: probably not a cluster member given its relatively blue colors; 3: Contaminated spectrum. Uncertainties on magnitudes are approximately 0.05-0.10 mag.}                                  
\end{center}                                                               
\end{table*}

Data reduction was performed with the ESO SINFONI pipeline version 3.0.0 under the ESO \emph{Reflex} environment version 2.8.5 \citep{freuding13}. Source spectra were extracted from data cubes using the QFitsView software version 3.1\footnote{QFitsView is developed by T. Ott at MPE and is available at \url{http://www.mpe.mpg.de/~ott/QFitsView/}.}. Atmospheric correction was made using standard stars observed just before or after science observations. To ensure optimal correction, we determined the radial velocity shift between the atmospheric absorption lines in the standard and target spectra. We used a $\chi^2$ analysis in the 2.06-2.08 $\mu$m wavelength range (rich in atmospheric features) for that purpose. The standard star spectrum was then shifted according to this radial velocity. The stellar Br$\gamma$ line of the standard stars was manually removed. We then divided the target spectrum by the standard spectrum to perform atmospheric correction. We finally normalized the targets' spectra by selecting manually continuum points that were subsequently used to define the continuum shape from a spline interpolation. The signal-to-noise ranges from $\gtrsim$100 for the brightest stars (e.g., star 12) to about 10 for the faintest ones.
In Fig.\ \ref{fig_chart} we have identified the 11 stars (objects 1-11) observed by \citet{chene13}. We have subsequently identified 40 additional stars in the SINFONI observations.

%%%%%%%%%%%%%%%%%%%%%%%%%%%%%%%%%%%%%%%%%%%%%%%%%%%%%%%%%%%%%%%%%%%%%%%%%%%%%%%%%%%%%%%%%%%%%%%%%%%%%%%%%%%%%%%%%%%%%%%%%%%%%%%
\section{Stellar content and spectral types}
\label{s_res}

\begin{figure*}[t]
\centering
\includegraphics[width=0.99\textwidth]{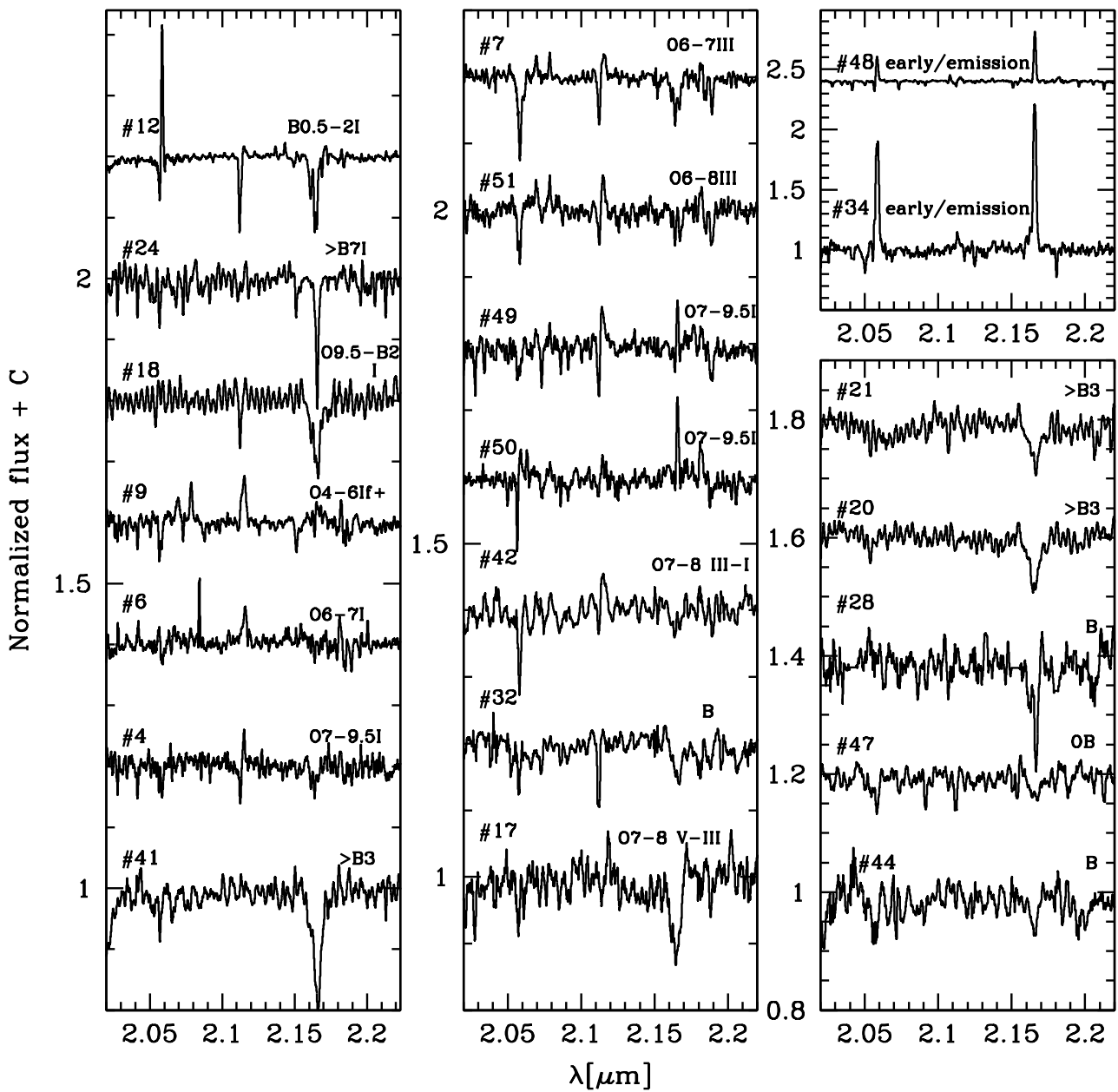}
\caption{Spectra of OB stars. The stars are ordered according to K-band magnitude; the brightest stars are on the top left and the faintest ones on the bottom right. The two emission line stars are shown in the top right panel.}
\label{specOB}
\end{figure*}

\begin{figure*}[t]
\centering
\includegraphics[width=0.99\textwidth]{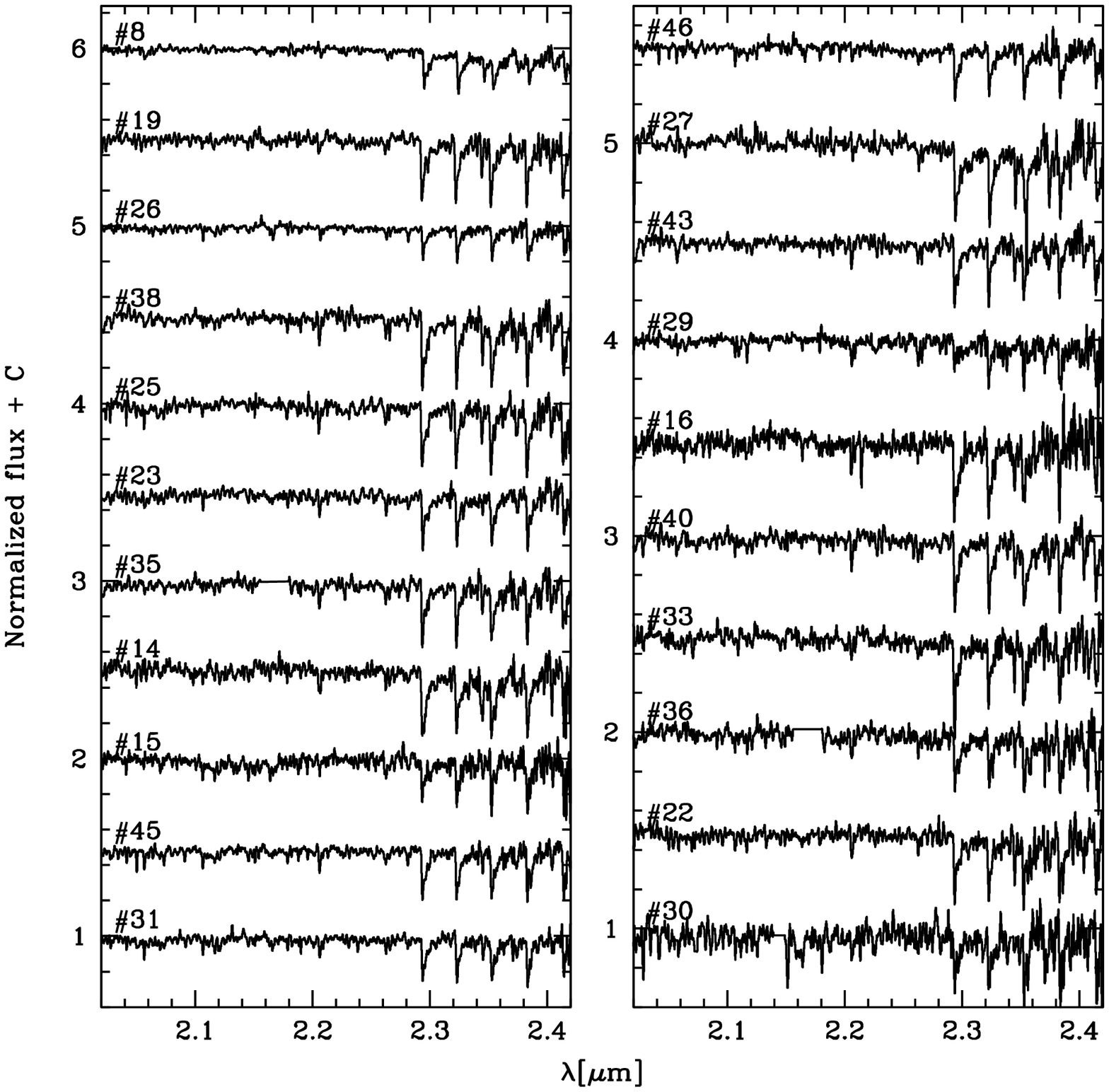}
\caption{Same as Fig.\ \ref{specOB} for stars showing CO absorption band-heads.}
\label{specCO}
\end{figure*}

The spectra were separated into those showing and not showing the strong CO absorption band-heads near 2.3 \mum. The latter all show \brg\ absorption or emission that are typical of OB and Wolf-Rayet stars. They are displayed in Figs.\ \ref{specOB}, \ref{fit_O}, and \ref{fit_wr}. CO absorption stars are shown in Fig.\ \ref{specCO} and are discussed in Sect.\ \ref{s_co}.

\begin{figure}[t]
\centering
\includegraphics[width=0.49\textwidth]{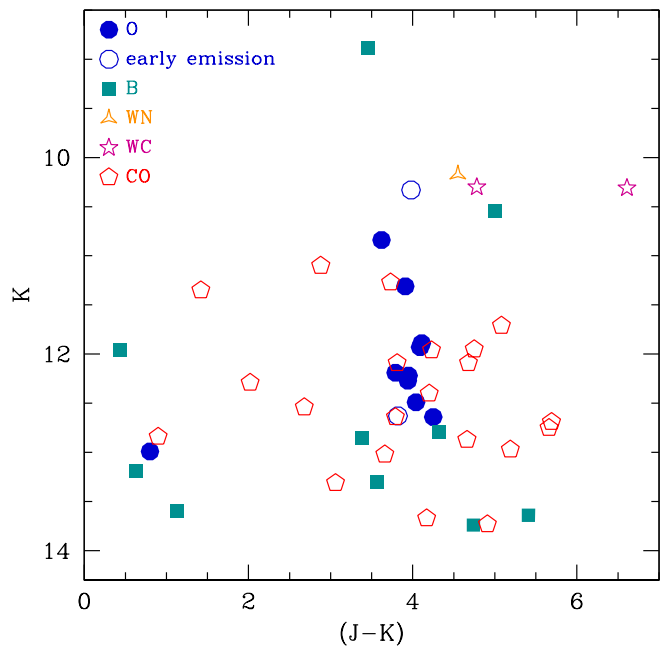}
\caption{K versus (J-K) color-magnitude diagram of the stars for which we obtained spectroscopy. Stars are symbol- and color-coded according to their estimated spectral type.}
\label{fig_cmd}
\end{figure}

We performed a spectral classification for OB and WR stars. For the former, we relied on comparison to the spectral atlases of \citet{hanson96,hanson05}. Our method is similar to that presented in \cite{martins07,martins08} and \cite{clark18}. O stars show \ion{He}{ii}~2.189 \mum\ while B stars do not. The strength of this helium line together with the strength and shape of the 2.112 \mum\ line complex (which includes \ion{He}{i}, \ion{C}{iii}, \ion{N}{iii,} and \ion{O}{iii} lines) discriminates between O subtypes. Stars earlier than O7 show \ion{C}{iv}~2.078 \mum\ emission. Luminosity classes are mainly assigned from the morphology of the Br$\gamma$ line. At spectral types earlier than O7, when moving from dwarfs to giants and supergiants, the line fills with wind emission and switches from absorption to emission. In the range O8-O9.7, Br$\gamma$ is narrower in supergiants and \ion{He}{i}~2.161 \mum\ becomes visible in the blue wing of Br$\gamma$. B stars' classification is difficult given the very limited number of spectral features in the K-band. In most cases, we could only assign a B spectral types from the absence of a \ion{He}{ii}~2.189 \mum\ line. For star 12, the presence of a weak emission around 2.11~\mum\ and the  strength of \ion{He}{i}~2.184 \mum\ correspond to B0.5-2 supergiants of the atlas of \citet{hanson05}. Finally, we stress that object 17 is a B-type star but given its peculiar colors (much bluer than the rest of the OB stars, see Fig.\ \ref{fig_cmd}) we suspect it is a foreground star. \citet{chene13} estimated that on average there should be 0.2$\pm$0.1 OB stars/arcmin in the direction of the clusters they studied (including VVV~CL074). The probability of having a foreground B-type star is thus not negligible.

The results of our spectral classification are shown in Table~\ref{tab_obs}. For most stars we could only constrain the spectral types and/or luminosity classes to a (narrow) range. This is due to the quality of our data and to the relatively low sensitivity of the diagnostic K-band lines to spectral type variations. Stars 4, 6, 7, and 9 are in common with \citet{chene13}. Stars 4 and 6 have a better signal-to-noise ratio in the present study and the spectral classification is thus revised compared to \citet{chene13}. Star 7 shows features typical of O6-7 stars (\ion{C}{iv} emission, \ion{He}{ii} present, absorption and emission  profile at 2.11 \mum) together with \brg\ in weak absorption. This is consistent with luminosity class III according to \citet{hanson96}.  For star 9 we revised the classification of \citet{chene13} from WN7/O4-6I+ to O4-6If+. A direct comparison of the star's spectrum to O4-6If+ stars from the Arches cluster \citep{martins08,clark18} revealed a very good match in relative and absolute line strengths. WN7 stars have stronger emission lines \citep{morris96}.

For WR stars, we compared the observed spectra to the atlases of \citet{figer97}. The presence of strong, relatively narrow (by WR standards) emission lines from \ion{He}{i} and hydrogen is characteristic of the WN8 spectral class. The presence of the \ion{N}{iii}~2.24 \mum\ emission doublet confirms this classification for star 3. The ratio of equivalent widths of \brg\ to \ion{He}{ii}~2.189 \mum\ (log(\brg/\ion{He}{ii}~2.189)=0.7) is typical of WN8(h) stars according to \citet{rc18}. Stars 2 and 13 are WR stars of class WC8 and WC9. They show broad \ion{C}{iii} and \ion{C}{iv} emission lines. They are mainly distinguished by the presence of \ion{He}{i} in WC9 stars. Lines are also broader in WC8 stars. The ratio of equivalent widths of \ion{C}{iii}~2.11 \mum\ to \ion{C}{iv}~2.189 \mum\ (log(\ion{C}{iii}~2.11/\ion{C}{iv}~2.189)=-0.22 and +0.18 for stars 2 and 13 respectively) are fully consistent with our classification (see Table 3 of \citet{rc18}).

Two stars (number 34 and 48) display narrow emission in \ion{He}{i}~2.058 \mum\ and \brg, and to a lower level in the line complex at 2.11 \mum. The full width at half maximum of \brg\ is 350 (250) \kms\ in stars 34 (48), favoring a stellar rather than a nebular origin. There is evidence of \ion{Si}{iv}~2.427 \mum\ emission in star 48. These emission features are typical of evolved massive stars. They are encountered in O and B emission stars (Oe and Be stars, sometimes in X-ray binary systems), B hypergiants, WN10-11h stars, and LBVs or candidate LBVs \citep[see][]{morris96,hanson96,clark18}. Distinction between these classes based solely on the K-band is difficult. \ion{Fe}{ii}~2.089 \mum\ and \ion{Na}{ii}~2.206-2.209 \mum\ are usually detected in emission in LBVs. Since we do not clearly identify these features, we exclude this class for stars 34 and 48. Other than that, we cannot unambiguously separate between OBe, B hypergiants, and WN10-11h stars. We thus attribute the classification type ``early/emission''.   

In total we identified 25 OB and WR stars, 19 being new discoveries. Figure\ \ref{fig_cmd} shows the K versus (J-K) color-magnitude diagram of the stars observed by SINFONI. The O stars define a relatively clear vertical sequence. The Wolf-Rayet stars, especially the WC stars, have on average redder colors than the OB stars. In addition to star 17, three additional objects classified as B or later than B (but not showing CO absorption) may be foreground stars given their blue color: stars 21, 28, and 41. The CO absorption stars show a wide range of colors. 

%%%%%%%%%%%%%%%%%%%%%%%%%%%%%%%%%%%%%%%%%%%%%%%%%%%%%%%%%%%%%%%%%%%%%%%%%%%%%%%%%%%%%%%%%%%%%%%%%%%%%%%%%%%%%%%%%%%%%%%%%%%%%%%
\section{Stellar parameters}
\label{s_param}

%%------------------------------------------------------------------
\subsection{OB and WR stars}
\label{s_obwr}

%%------------------------------------------------------------------
\subsubsection{Extinction and distance}

To estimate the amount of extinction in the K band ($A_{K}$), we used the color excess of the three O supergiants with the best spectra: stars 4, 6, and 7. For each star, we adopted the absolute magnitudes and colors of \citet{mp06}. Our results thus depend on the reliability of these calibrations.

In Fig.\ \ref{fig_cmd}, stars 4, 6, and 7 are the blue circles with K-band magnitudes around 12. Their colors are representative of the brightest members of the cluster. Hence they are well suited to define an average extinction. 
The choice of the extinction law is critical. To minimize its effect on the results, we computed the average value of $A_{J}/A_{K}$ from Table 5 of \citet{wang14}, which compiles results from ten different studies of the extinction law. We obtained $A_{J}/A_{K}$=2.85$\pm$0.48. We then estimated $A_{K} = \frac{1}{A_{J}/A_{K}-1} \times$ E(J-K) for each star, using the intrinsic colors of \citet{mp06}, leading to  $A_{K} =$ 2.27$\pm$0.34. We subsequently calculated the distance modulus from $A_{K}$, the absolute and observed magnitudes, and the associated errors. We finally adopted the average distance of the three stars as the cluster's distance, which is found to be 10.2$\pm$1.6 kpc. Given the cluster's galactic coordinates (\textit{l}=336.37$^{o}$, \textit{b}=0.19$^{o}$), this distance places VVV~CL074 close to the intersection of the Norma arm and the galactic bar. This region already hosts the young massive clusters Mercer 81 \citep{davies12a} and VVV~CL086 \citep{ram14b}.  [DBS2003]-179 \citep{borissova12} is also relatively nearby, though closer to the galactic center. VVV~CL074 is thus one of the farthest young massive clusters uncovered so far.

%%------------------------------------------------------------------
\subsubsection{Stellar parameters}

For O and WR stars we used the atmosphere code \emph{CMGFEN} to produce synthetic spectra. The full description of the code can be found in \citet{hm98}. We adopted solar abundances from \citet{ga10} in our computations, which included H, He, C, N, O, Ne, Mg, Si, S, Ar, Ca, Fe, and Ni ions. We have assumed He/H=0.1 for all stars. We refer the reader to \citet{herve16} for further information on the atmosphere models and synthetic spectra. The synthetic spectra were subsequently compared to observed data in order to determine the effective temperature and surface abundances (for WR stars).

\begin{figure*}[t]
\centering
\includegraphics[width=0.32\textwidth]{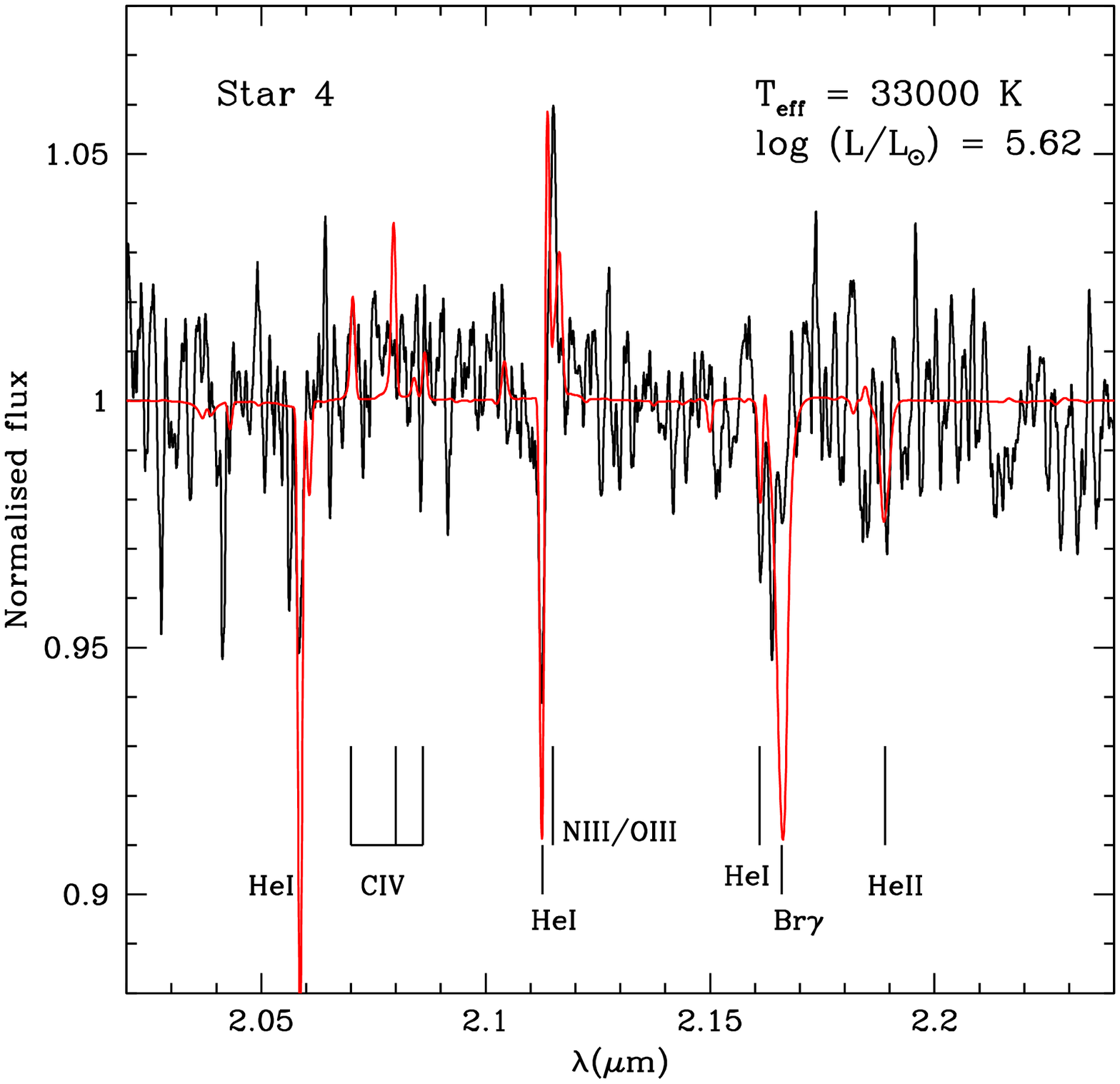}
\includegraphics[width=0.32\textwidth]{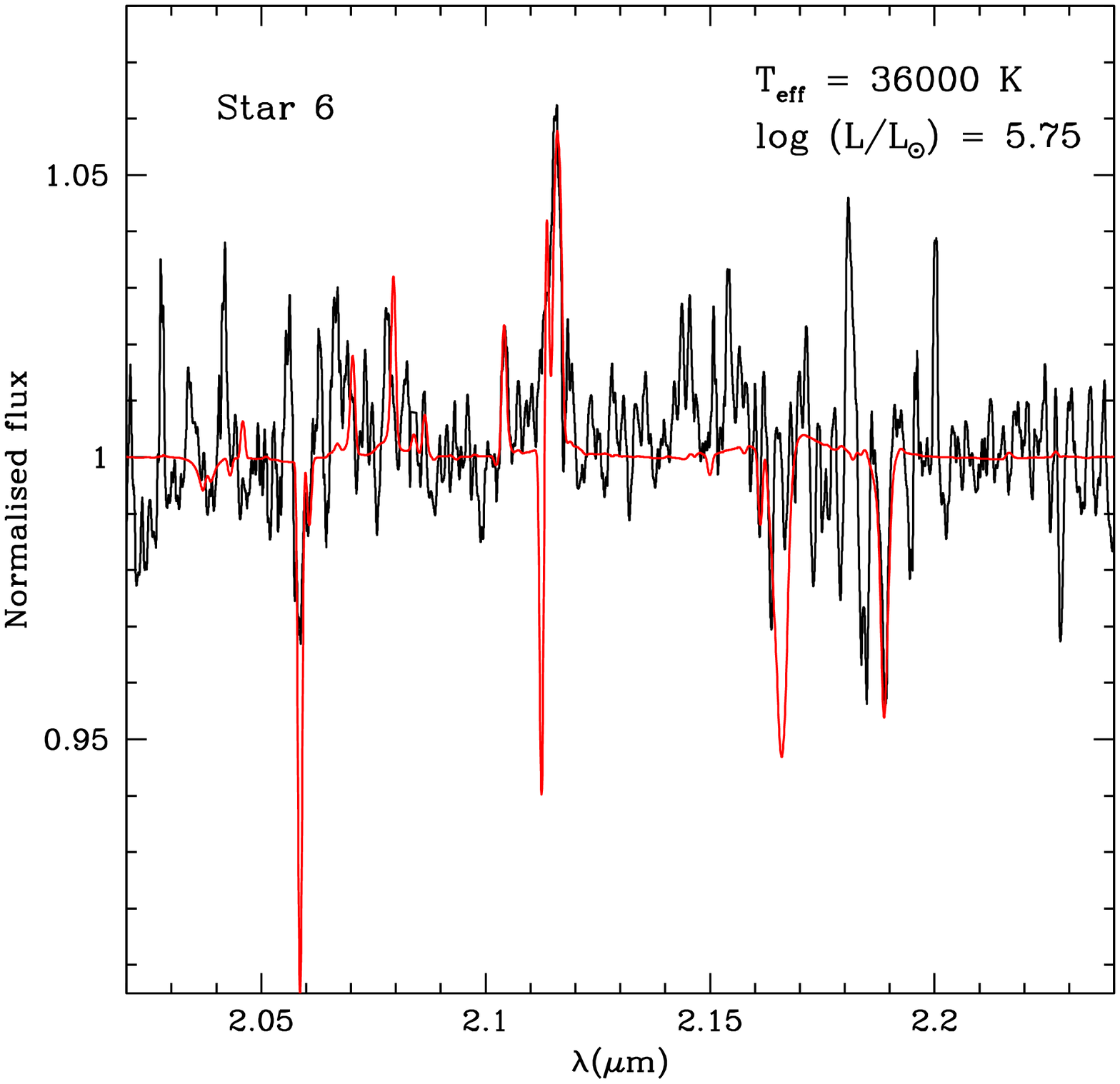}
\includegraphics[width=0.32\textwidth]{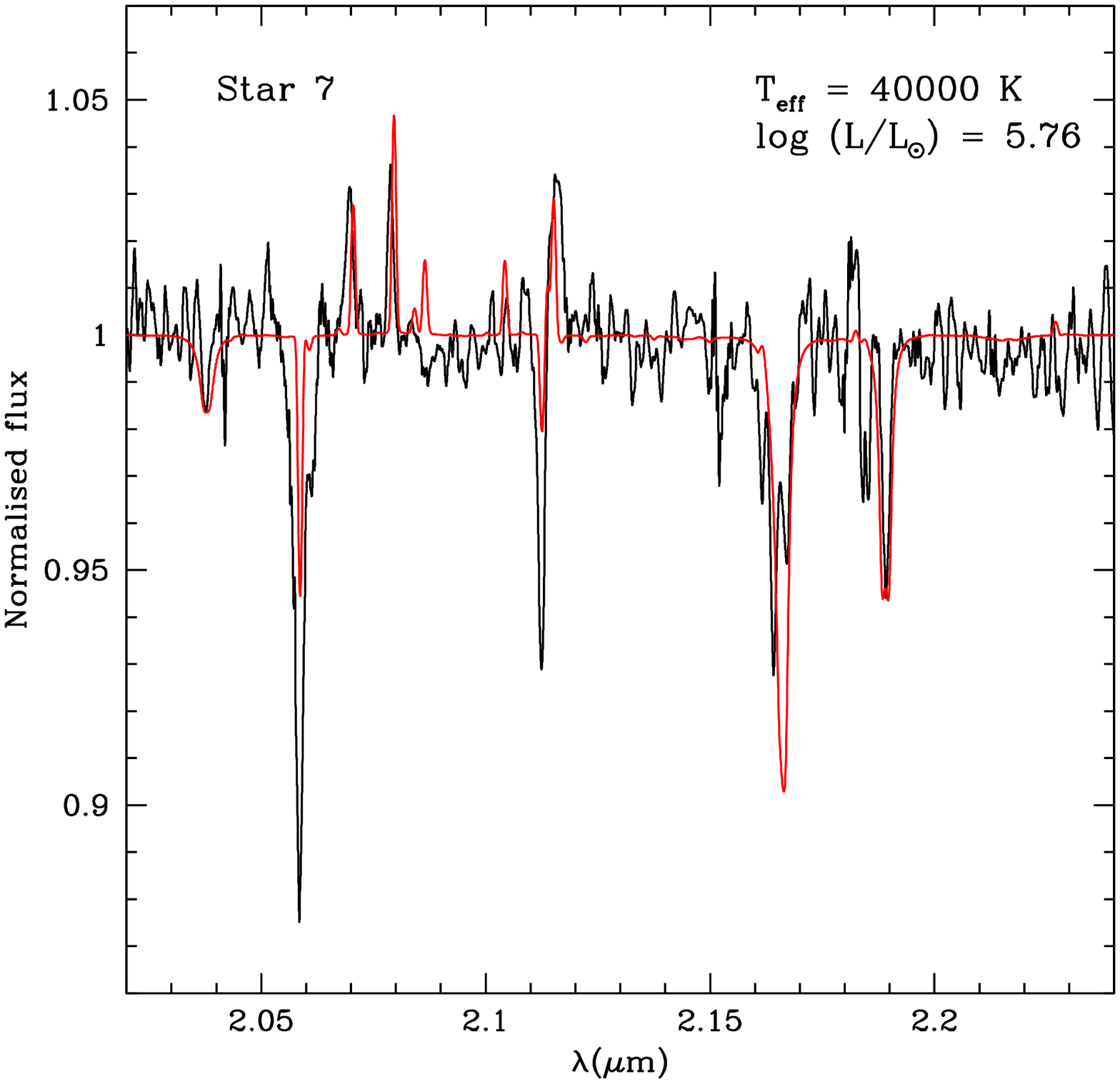}\\
\includegraphics[width=0.32\textwidth]{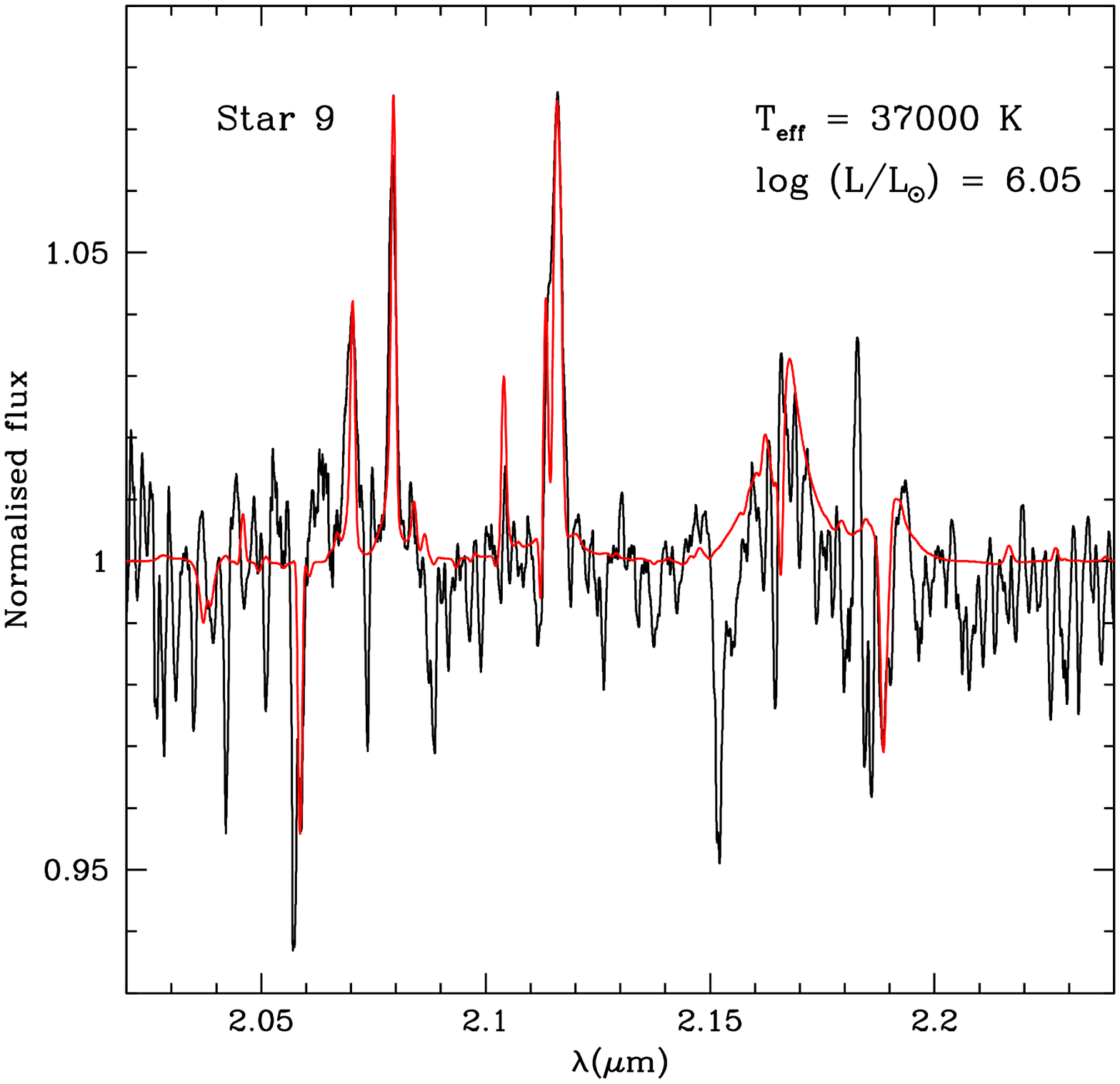}
\includegraphics[width=0.32\textwidth]{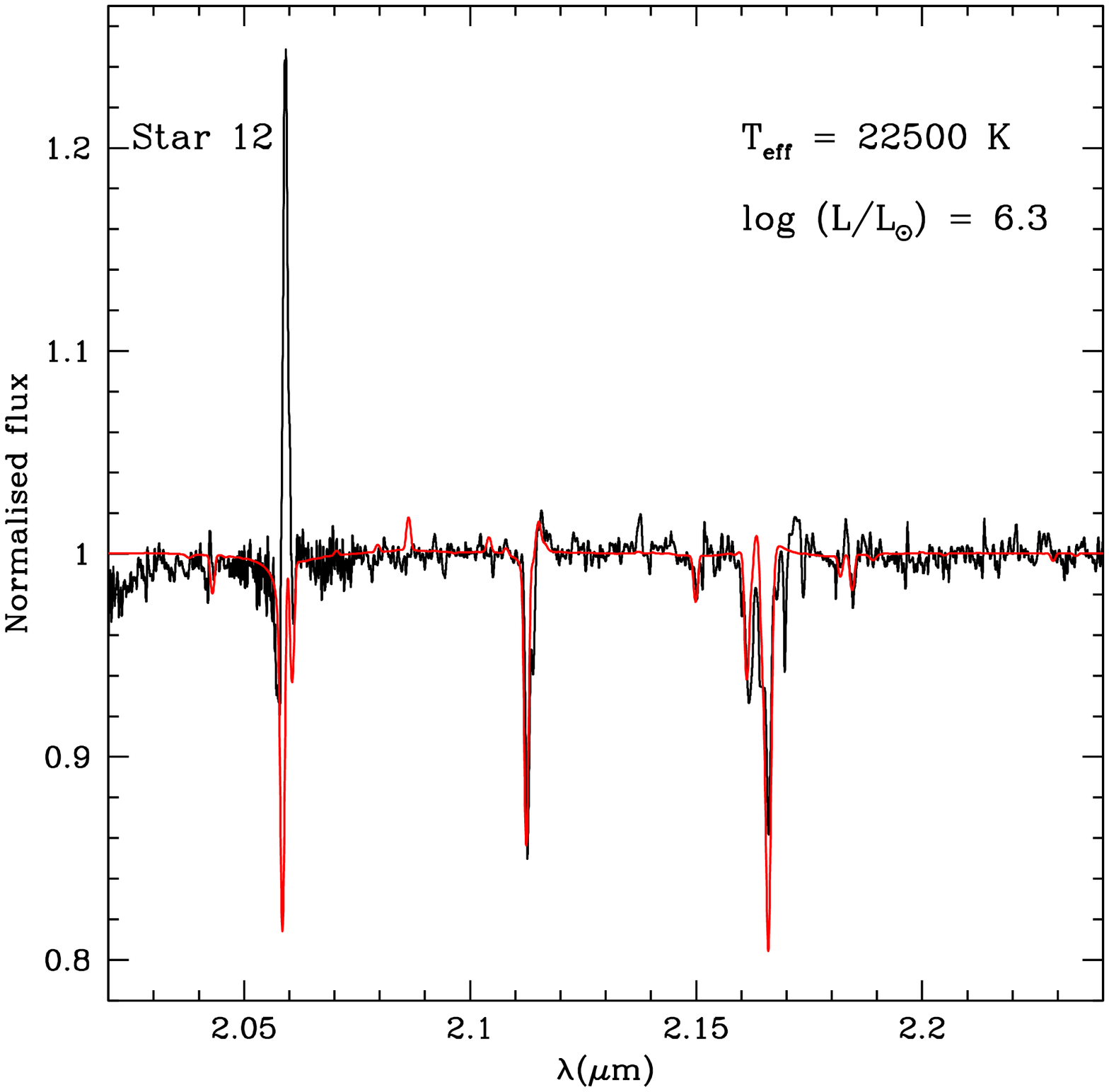}
\includegraphics[width=0.32\textwidth]{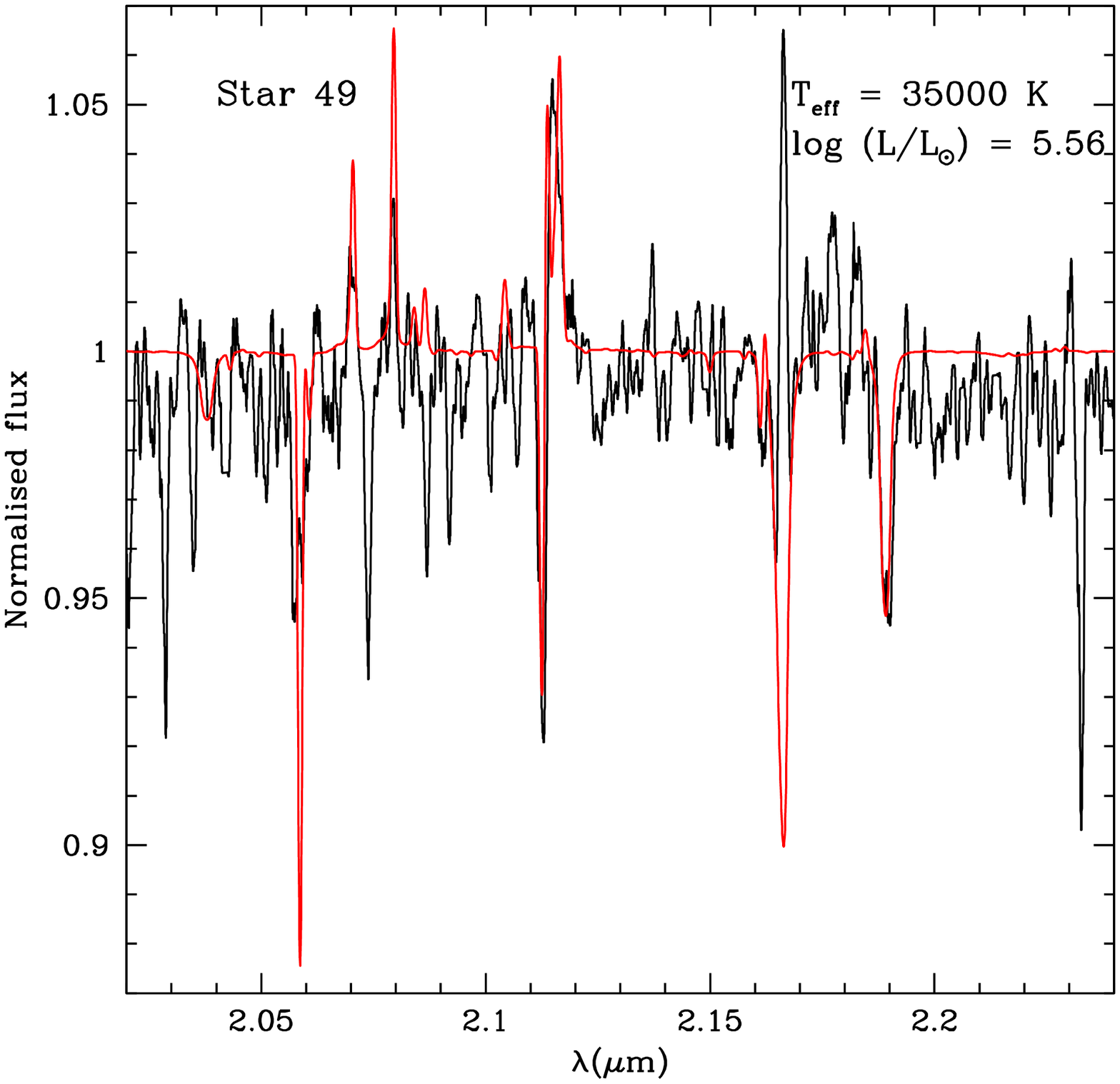}\\
\includegraphics[width=0.32\textwidth]{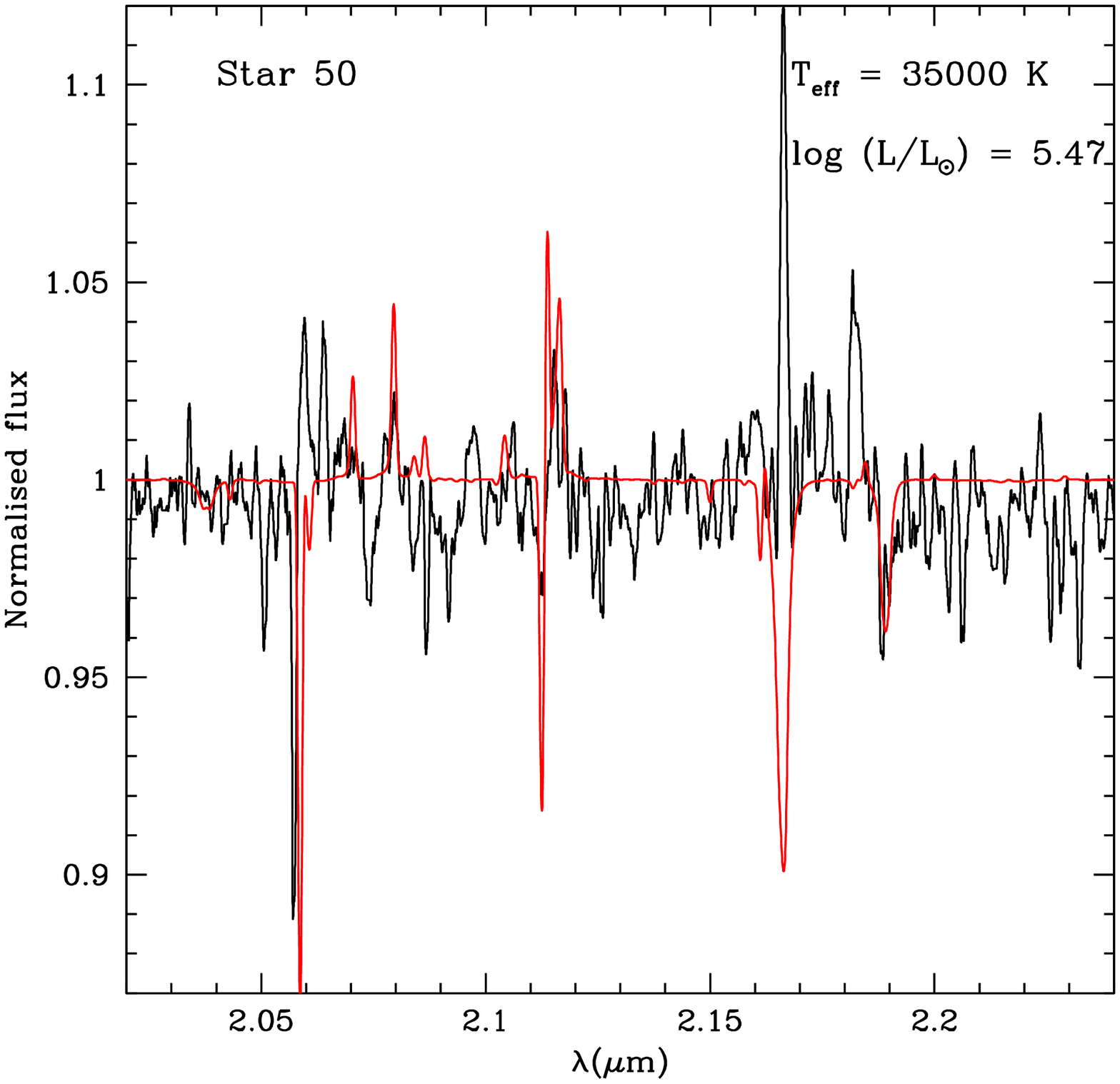}
\includegraphics[width=0.32\textwidth]{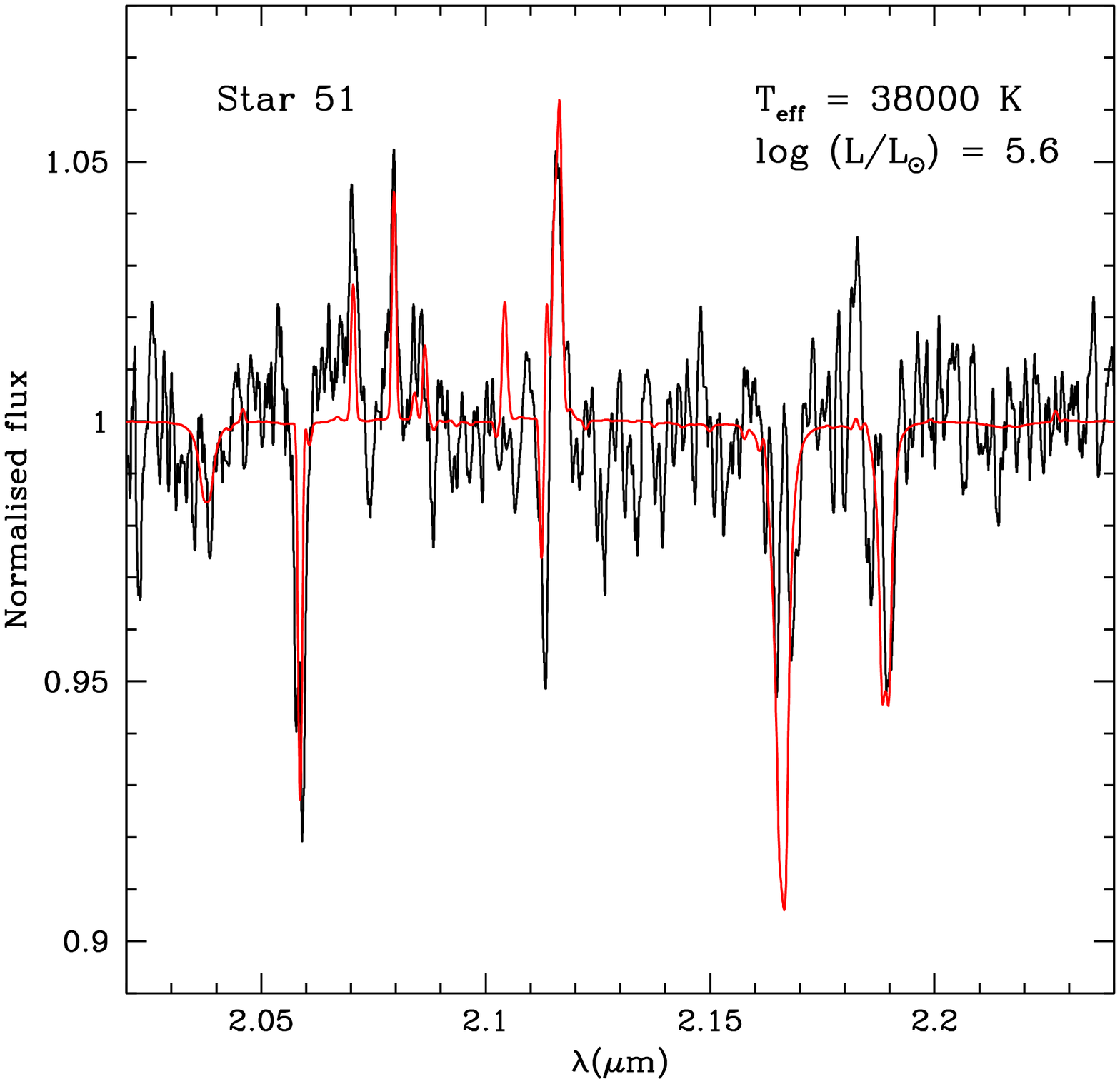}\\
\includegraphics[width=0.32\textwidth]{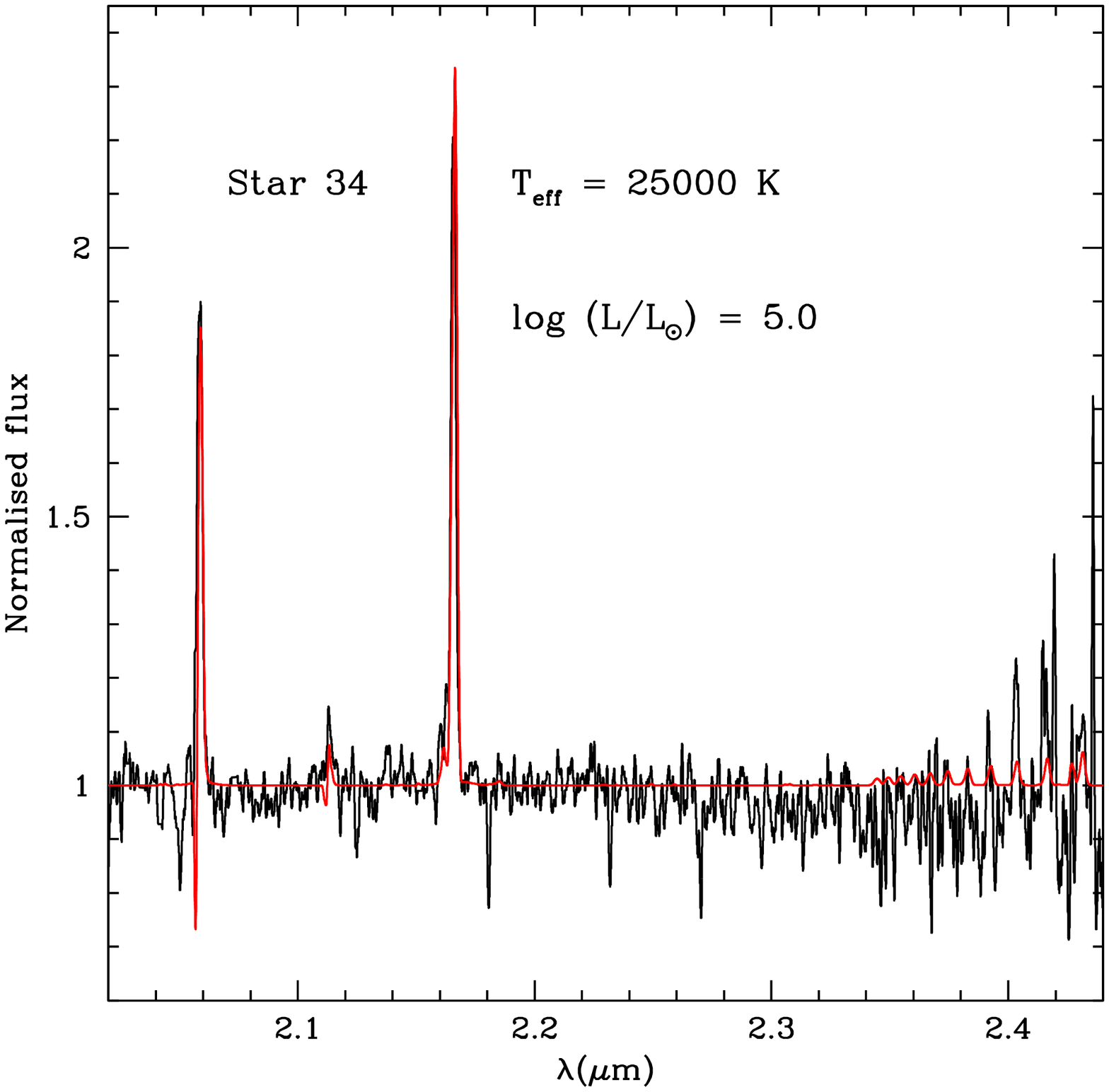}
\includegraphics[width=0.32\textwidth]{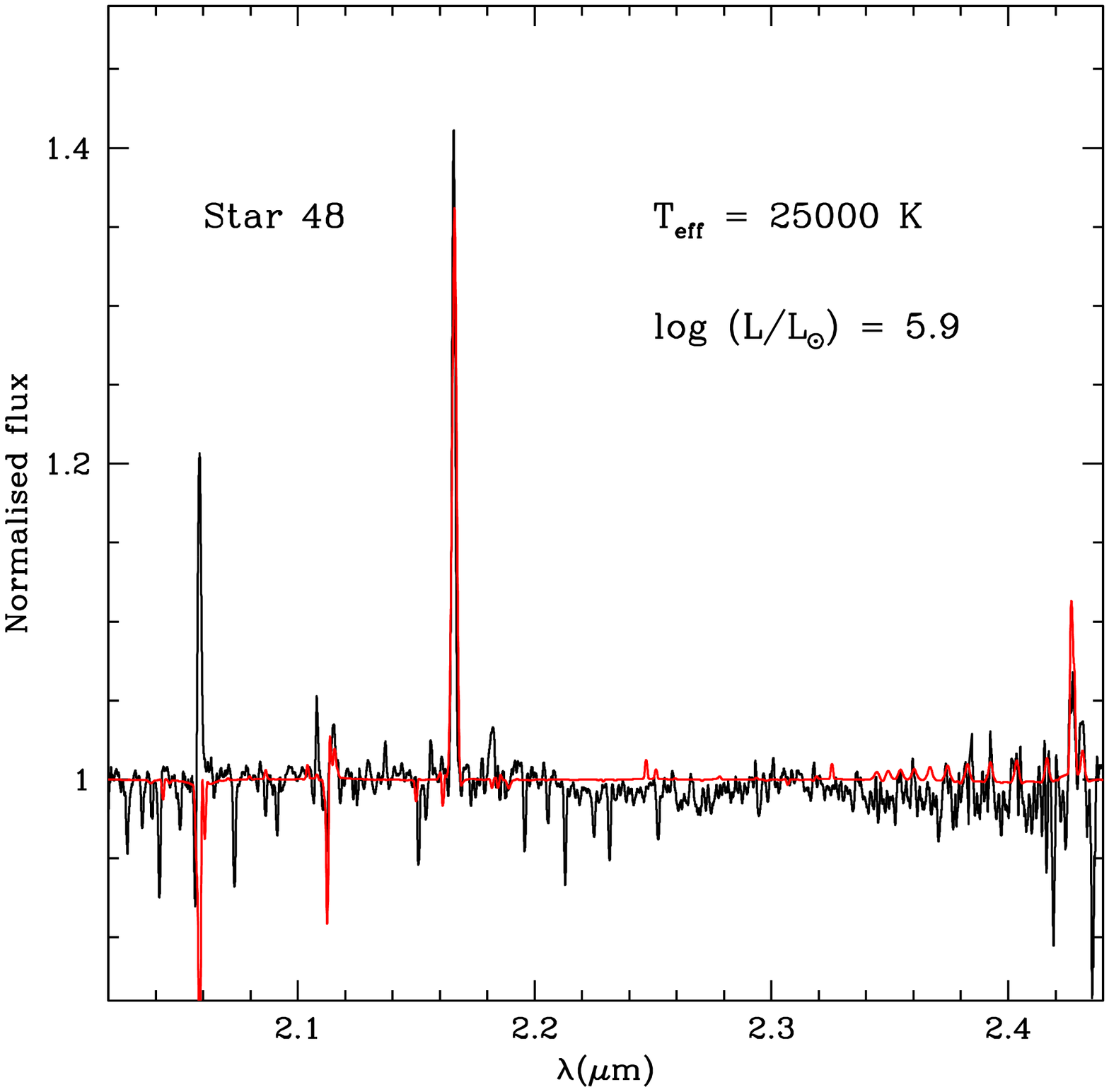}
\caption{Best fit (red) of the observed spectrum (black) of stars for which the spectrum quality was sufficient to perform a spectroscopic analysis. The main diagnostic lines are indicated in the upper left panel.}
\label{fit_O}
\end{figure*}

For O stars the strength of \ion{He}{ii}~2.189 \mum\ was the main \teff\ indicator. When the quality of the spectrum was too low and/or there was no sign of \ion{He}{ii}~2.189 \mum, \teff\ was assigned from spectral type using the calibrations of \citet{msh05}, \citet{paul06}, and \citet{nieva13}. This applies to stars 17, 18, and 42. The stellar luminosity was calculated from \teff, the observed K-band magnitude,  K-band bolometric corrections \citep{mp06}, and the distance and extinction estimated above. The best fits are shown in Fig.\ \ref{fit_O} and the stellar parameters are summarized in Table \ref{tab_param_O}. They include stars 34 and 48 for which the effective temperature was broadly constrained from the absence of \ion{He}{ii}~2.189 \mum, the shape of the 2.11~\mum\ emission complex, and the strength of \ion{Si}{iv}~2.427~\mum. For these stars, the narrow emission in \ion{He}{i}~2.058~\mum\ and \brg\ could be reproduced with a terminal wind velocity of 300 \kms\ and mass loss rates on the order of 1 to 3 $10^{-6}$ \myr. For the B supergiant star 12, we used an average of the effective temperatures determined by \citet{paul06} for B0.5 to B2 stars. We stress that all our \teff\ estimates assume a solar helium content.

\begin{figure*}[t]
\centering
\includegraphics[width=0.32\textwidth]{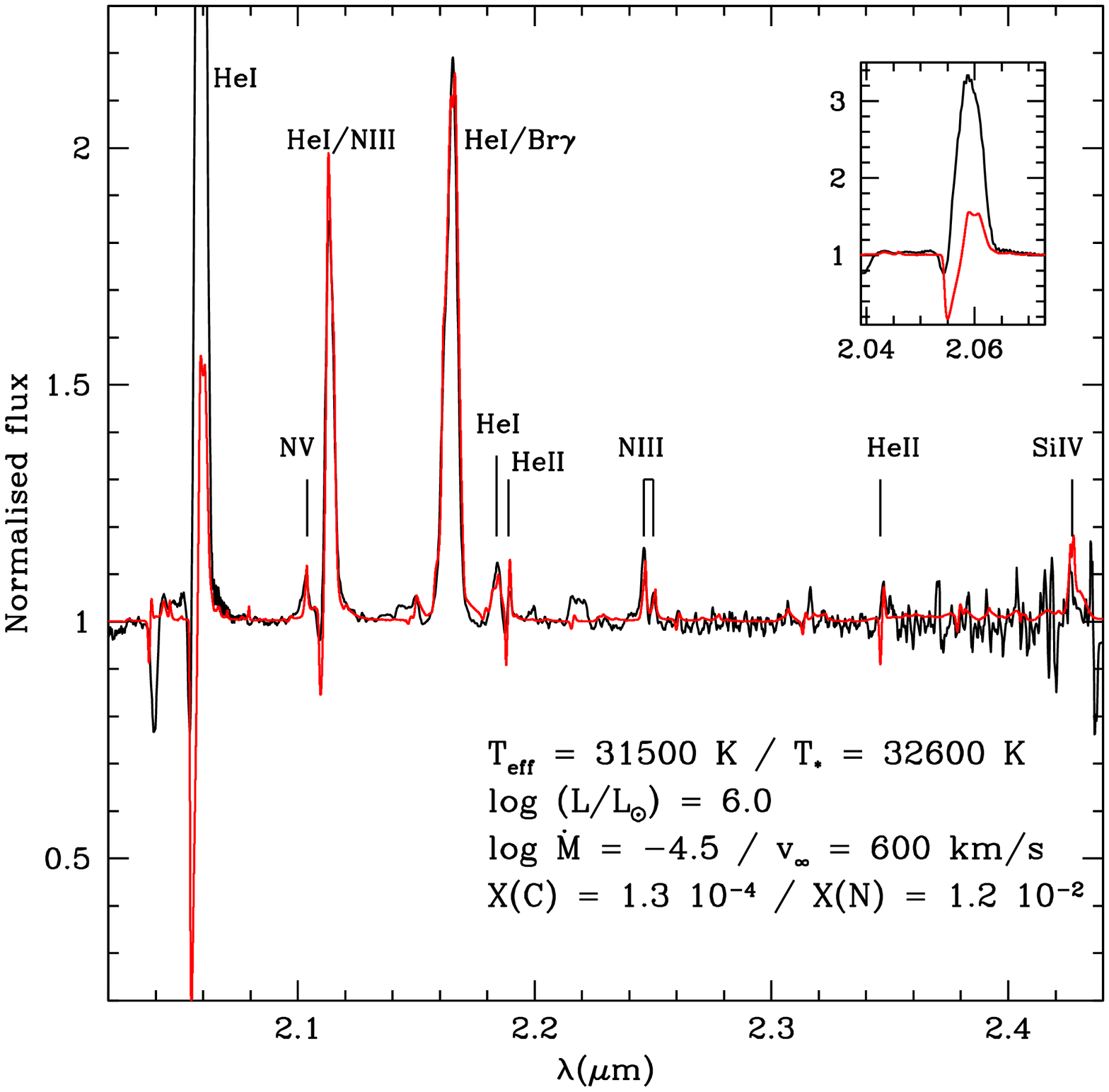}
\includegraphics[width=0.32\textwidth]{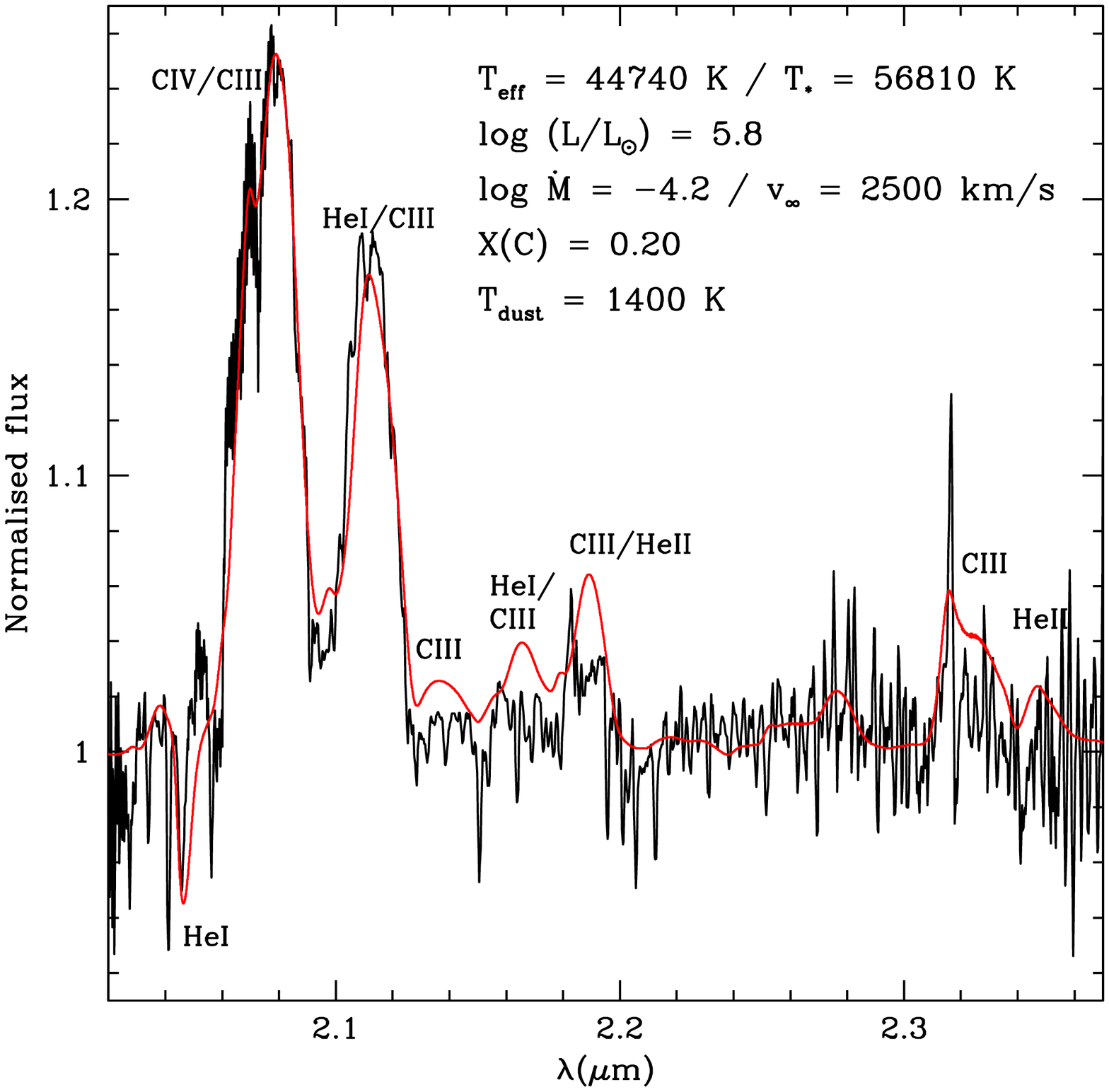}
\includegraphics[width=0.32\textwidth]{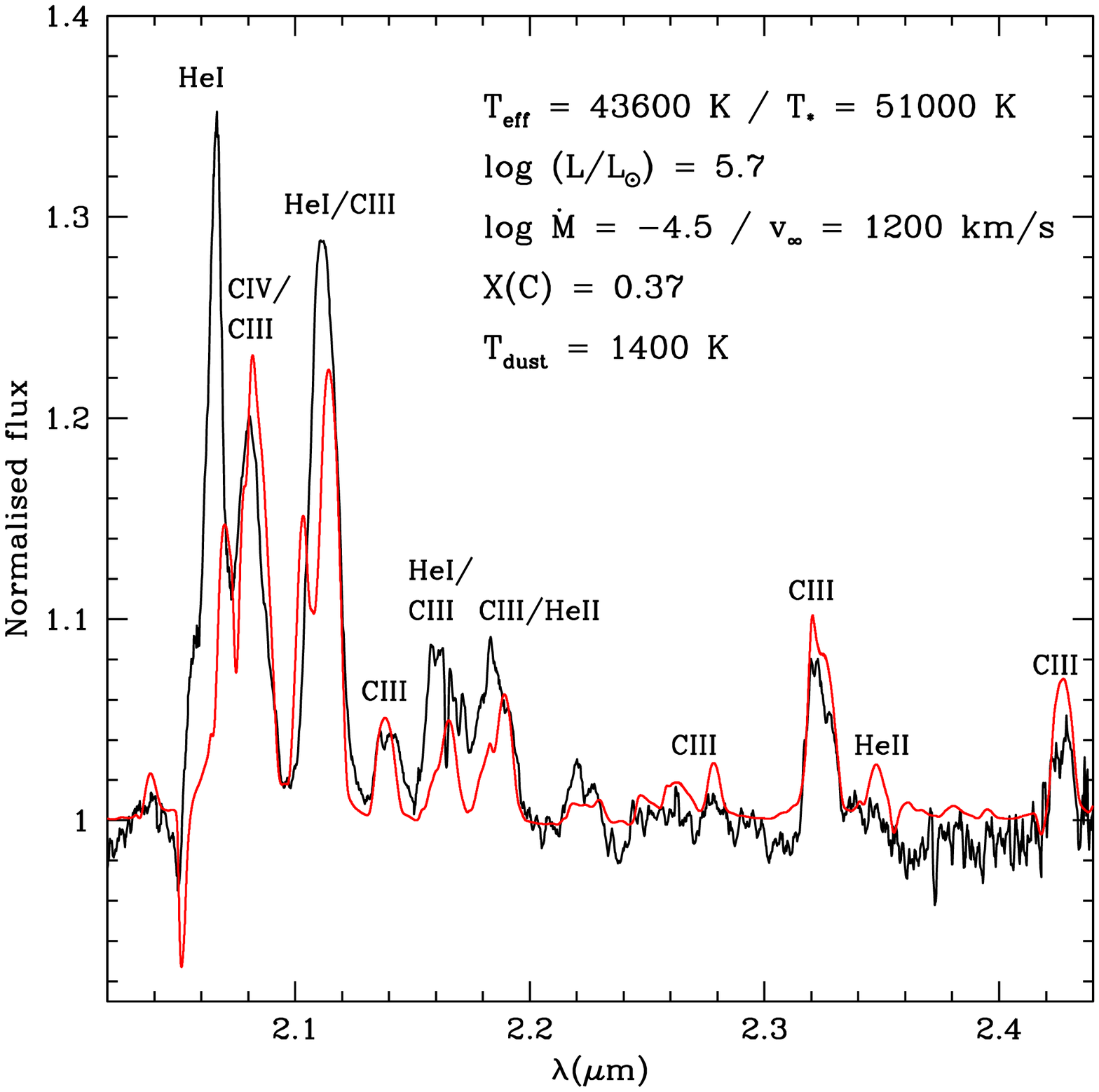}
\caption{Best fit (red) of the observed spectrum (black) of the three Wolf-Rayet stars: stars 3 (WN8, left), 2 (WC8, middle), and 13 (WC9, right). The main lines are indicated in each panel.}
\label{fit_wr}
\end{figure*}

To determine the effective temperature of the WN8 star number 3 we relied on the relative strength of \ion{He}{i} and \ion{He}{ii} lines. The mass loss rate and the helium to hydrogen ratios were determined from the absolute strength of helium and hydrogen lines. The nitrogen content was constrained from the \ion{N}{iii} lines at 2.24 \mum. Finally the upper limit on the carbon abundance was estimated from the absence of \ion{C}{iv} lines between 2.05 and 2.08 \mum. The luminosity was adjusted so that the absolute K-band magnitude of the model matches the observed value (after taking into account the extinction and distance modulus). The stellar parameters are very similar to those obtained by \citet{herve16}. The imperfect fit of the \ion{He}{i}~2.06 \mum\ line (see Fig.\ \ref{fit_wr}, left panel) is a well-known effect of the extreme sensitivity of this feature to details of the modeling \citep[atomic data, microturbulence, line-blanketing; see, e.g.,][]{paco06}.

For the two WC stars, we used the same method as for the WN8 object. We assumed that no hydrogen was observed at the surface of the star. We noticed that the emission level of most lines is relatively low compared to normal WC stars \citep{figer97}. This is usually explained by the presence of dust continuum emission around the star \citep[e.g.,][]{crowther06}. We thus followed the standard procedure that consists in adding a 1400 K blackbody contribution to the stellar emission \citep{rosslowe}. The level of dust emission was adjusted so the stellar lines are diluted enough to reproduce their observed emission level. Photometry was computed from the resulting total spectral energy distribution and compared to absolute magnitudes in order to determine the stellar luminosity. This process ensures a determination of the relative stellar and dust flux emission, as well as of the absolute stellar luminosity. In practice, dust contributed to 80\% of the K-band flux for both stars. We found out that it was never possible to obtain a good fit of all spectral lines at the same time. The spectrum is dominated by helium and carbon lines. The effective temperature was mainly set from the ratio of \ion{C}{iv} to \ion{C}{iii} lines. The mass loss rate, helium, and carbon abundances were obtained from the emission line level, modulated by the dust contribution. The best fit shown in Fig.\ \ref{fit_wr} should be viewed as indicative of average stellar parameters given the uncertainties on continuum normalization, dust properties,, and atmospheric structure (e.g. shape of the velocity law). We note, however, that our stellar parameters, gathered in Table \ref{tab_param_WR}, are consistent with those of similar objects \citep{paul07,williams15,sander18}. The WC9 star is rather luminous, but still in the luminosity range advocated by \citet{sander18} for galactic stars with known \emph{Gaia} distances. We caution that the luminosity of the two WC stars may be overestimated. The presence of dust is usually associated to a wind-colliding region in a binary system \citep{williams01,tut06,tut08}. If a secondary component is present, it may contribute to the observed K-band magnitude. \citet{najarro17} included the emission of a companion O star in their modeling of a dusty WC star in the Quintuplet cluster (see also below). They relied on a wide spectral range extending to the J-band, where the contribution of a companion OB star is expected to be stronger relative to the WC star. In our case, the limited spectral coverage does not allow such a modeling, unless assuming arbitrary parameters for the companion. We thus assumed that all the K-band emission was due to both the WC stars and the dusty region. Consequently, the stellar luminosities are upper limits.

\begin{table}
\begin{center}
\caption{Parameters of the OB stars.}
\label{tab_param_O}
\begin{tabular}{lcccccccc}
\hline
Star        & Spectral & Teff  & \lL &   \\    
            &  type    & [kK]  &     &   \\
\hline
\smallskip
4           & O7-9.5 I       & 33.0$\pm$2.0  & 5.62$\pm$0.23 &  \\
6           & O6-7 I         & 36.0$\pm$2.0  & 5.75$\pm$0.20 &  \\
7           & O6-7 III       & 40.0$\pm$2.0  & 5.76$\pm$0.21 &  \\
9           & O4-6 If+       & 37.0$\pm$2.0  & 6.01$\pm$0.20 &  \\
12          & B0.5-2 I       & 22.5$\pm$4.0  & 6.36$\pm$0.30 &  \\ 
17          & O7-8 V-III     & 34.0*$\pm$3.0  & 5.24$\pm$0.21 &  \\
18          & O9.5-B2 I      & 26.0*$\pm$6.0  & 5.76$\pm$0.35 &  \\
34          & early/emission & 25.0$\pm$3.0  & 4.99$\pm$0.24 &  \\
42          & O7-8III-I      & 34.0*$\pm$3.0 & 5.38$\pm$0.22 &  \\
48          & early/emission & 25.0$\pm$3.0  & 5.91$\pm$0.24 &  \\
49          & O7-9.5 I       & 35.0$\pm$2.0  & 5.56$\pm$0.22 &  \\
50          & O7-9.5 I       & 35.0$\pm$2.0  & 5.47$\pm$0.23 &  \\ 
51          & O6-8 III       & 38.0$\pm$2.0  & 5.68$\pm$0.22 &  \\
\hline
\end{tabular}
\tablefoot{Adopted parameters are marked with an asterisk (*) symbol.}
\end{center}
\end{table}

\begin{table*}
\begin{center}
\caption{Parameters of the WR stars. He/H is the number ratio, while X(C) and X(N) are mass fractions.}
\label{tab_param_WR}
\begin{tabular}{lcccccccccc}
\hline
Star        & Spectral & T$_{eff}$  & T$_{*}$  &   logg & \lL & log $\dot{M}$ & \vinf\ &  He/H  &  X(C)  &   X(N)     \\    
            &  type    & [kK]    & [kK]  &            &     &               & \kms\  &      &        &    \\
\hline
\smallskip
3           & WN8      & 31.5$\pm$3.0      & 32.6$\pm$3.0     &  3.5* & 6.0$\pm$0.25 & -4.5 & 600*  & 1.7   & $<$1.3 10$^{-4}$  & 1.2 10$^{-2}$ \\
2           & WC8      & 44.1$\pm$5.0      & 49.3$\pm$5.0     &  4.0* & $<$5.8$\pm$0.25 & -4.5 & 2200 & no H*  & 0.11          & --          \\
13          & WC9      & 44.0$\pm$5.0      & 51.0$\pm$5.0     &  4.7* & $<$5.7$\pm$0.25 & -4.5 & 1200* & no H*  & 0.37          & --          \\
\hline
\end{tabular}
\tablefoot{Adopted parameters are marked with an asterisk (*) symbol. Uncertainties on surface abundances are approximately 50\%.}
\end{center}
\end{table*}

%%------------------------------------------------------------------
\subsection{CO absorption stars}
\label{s_co}

Stars showing CO absorption band-heads, if they belong to the VVV CL074 cluster, can be red supergiants (RSG) or asymptotic giant branch (AGB) stars depending on their initial mass. Main sequence stars with CO band-heads are too faint and have not been detected in our observations, unless they are foreground stars. RSGs are evolved massive stars, while AGB stars are the descendents of lower mass stars. RSGs are observed after about 10 Myr in young clusters \citep{davies07,davies08}. If present, they may thus be part of the young massive stars population of the cluster.

Galactic RSGs have bolometric magnitudes between
-9.0 and -4.0 \citep{levesque05,clark09,messineo14}. Thus their luminosities are such that \lL > 3.6. To check if RSG stars are present among our CO absorption stars, we have estimated their luminosities in the following way. First, we determined their effective temperature from the CO index defined by \citet{blum96}. The index is 100$\times (1-\frac{F_{cont}}{F_{band}}),$ where $F_{cont}$ and $F_{band}$ are the fluxes in two wavelength ranges centered at 2.284 and 2.302 \mum,\ respectively. The two ranges are 0.015 \mum\ wide. Once calculated for all our CO stars, the index is compared to the relation CO index - \teff\ defined by \citet{maness07} to estimate \teff\ for each star. A bolometric correction is subsequently calculated from Eq.~4 of \citet{buzzoni10}. Assuming an extinction A$_{K}$ = 2.27 (see above) and a distance of 10.2 kpc, we determined the K-band absolute magnitudes, which were combined with the bolometric corrections to yield the luminosities. The results are gathered in Table \ref{tab_CO}. We see that all stars have luminosities lower than about 3.4, except stars 8 and 26 that are just at the limit of red supergiants (\lL $\sim$ 3.6). We thus consider most CO absorption stars as foreground objects. Stars 8 and 26 are also most likely foreground evolved giant stars although we cannot firmly exclude, from a purely observational point of view, that they are cluster members. If this was true, both stars would have to be older than the bulk of massive stars, implying multiple episodes of star formation in VVV~CL074. This is not favored in view of the coevality observed in most young massive clusters. 

To further test the possibility that the CO absorption stars are cluster member, we attempted to determine radial velocities for both OB and CO absorption stars. We used synthetic spectra taken from the \emph{POLLUX} database\footnote{\url{http://npollux.lupm.univ-montp2.fr/}} \citep{pollux} and performed cross-correlations with the observed spectra. We obtain an average velocity of -30$\pm$61 \kms\ (+44$\pm$58 \kms) for the OB (CO absorption) stars. The difference between both groups is thus barely significant at the 1$\sigma$ level. We conclude that radial velocities do not lead to strong conclusions regarding the cluster membership of the CO absorption stars.

\begin{table}
\begin{center}
\caption{Estimated parameters of the CO absorption stars.}
\label{tab_CO}
\begin{tabular}{lccc}
\hline
Star& MK   & Teff  &  \lL  \\    
    &      & [K]   &       \\
\hline
\smallskip
 8  & -6.21 & 4871  & 3.60 \\
14  & -5.22 & 4208  & 3.03 \\
15  & -5.02 & 4770  & 3.10 \\
16  & -4.44 & 4136  & 2.70 \\
19  & -6.04 & 4338  & 3.40 \\
22  & -3.64 & 4482  & 2.48 \\
23  & -5.35 & 4986  & 3.28 \\
25  & -5.36 & 4669  & 3.21 \\
26  & -5.96 & 5447  & 3.63 \\
27  & -4.62 & 4078  & 2.76 \\
29  & -4.47 & 4972  & 2.93 \\
30  & -3.58 & 4655  & 2.50 \\
31  & -4.77 & 5044  & 3.06 \\
33  & -4.29 & 4568  & 2.76 \\
35  & -5.22 & 4655  & 3.15 \\
36  & -4.00 & 4712  & 2.68 \\
38  & -5.60 & 4208  & 3.19 \\
40  & -4.34 & 4669  & 2.80 \\
43  & -4.56 & 4583  & 2.87 \\
45  & -4.91 & 4900  & 3.09 \\
46  & -4.67 & 5058  & 3.03 \\
\hline
\end{tabular}
\end{center}
\end{table}

%%%%%%%%%%%%%%%%%%%%%%%%%%%%%%%%%%%%%%%%%%%%%%%%%%%%%%%%%%%%%%%%%%%%%%%%%%%%%%%%%%%%%%%%%%%%%%%%%%%%%%%%%%%%%%%%%%%%%%%%%%%%%%%
%%%%%%%%%%%%%%%%%%%%%%%%%%%%%%%%%%%%%%%%%%%%%%%%%%%%%%%%%%%%%%%%%%%%%%%%%%%%%%%%%%%%%%%%%%%%%%%%%%%%%%%%%%%%%%%%%%%%%%%%%%%%%%%
\section{Discussion}
\label{s_disc}

%%------------------------------------------------------------------
\subsection{Evolutionary status}
\label{s_evolstat}

\begin{figure}[t]
\centering
\includegraphics[width=0.49\textwidth]{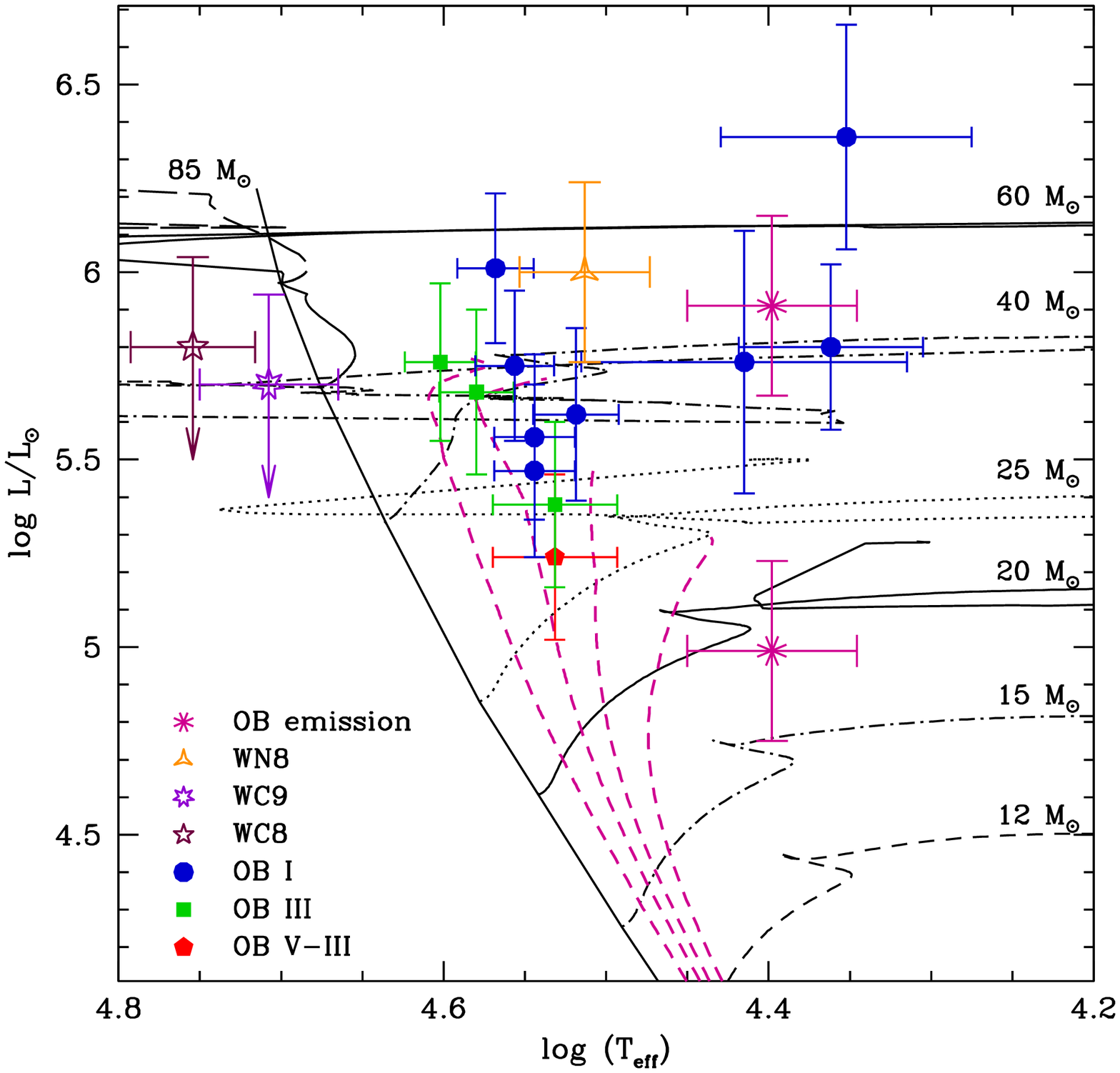}
\caption{HR diagram. The evolutionary tracks including rotation are from \citet{ek12}. Different symbols stand for the sample stars and refer to different luminosity classes depending on shape and color. The dashed purple lines are the 3, 5, 6, and 8 Myr isochrones cut to avoid the post-main sequence phases for clarity. }
\label{figHR}
\end{figure}

Figure\ \ref{figHR} shows the HR diagram of the sample stars with the evolutionary tracks and isochrones of \citet{ek12}. Most of the OB stars are located between the 3 and 6 Myr isochrones. They are located on the main sequence and have initial masses between $\sim$20 and $\sim$60 \msun. The two WC stars are located on the left side of the zero-age main sequence (ZAMS). According to the tracks of \citet{ek12}, they could be the descendants of stars with initial masses around 40-60 \msun. The WN8 star could have a progenitor with a mass in the same range. Its position on the right side of the ZAMS would be consistent with an earlier evolutionary state compared to the WC stars. The global picture resulting from these considerations would be that VVV CL074 is a 3-6 Myr old cluster hosting a population of main sequence OB stars with masses lower than 40-60 \msun, a mass that would correspond to that of the turn-off. Three stars of WR type could be the descendants of stars slightly more massive than this turn-off mass.

Star 12 is the most luminous object of the cluster. It is a B supergiant and its position in the HR diagram is consistent with a relatively massive star (M$\gtrsim$ 60 \msun) that has evolved off the main sequence. Given the large uncertainty on its effective temperature, and thus luminosity, it can be regarded as a star with initial mass just above the cluster turn-off that is evolving towards the red part of the HR diagram. We also stress that if it were a binary star, its luminosity would be slightly overestimated, explaining its position in the HRD. Under the binary hypothesis, it could also be overluminous because of mass transfer. Under such circumstances, the mass gainer experiences a sudden luminosity increase \citep{wellstein01,langer12}.

Opposite to star 12, star 34 is the least luminous of the sample. It is located much farther away from the ZAMS than the bulk of OB stars. Star 34 has a spectrum very similar to star 48 which, according to its position in the HR diagram, makes it an evolved O supergiant with a dense wind, close to the Wolf-Rayet phase. The relatively low luminosity of star 34 may be explained by an episode of mass transfer in a binary system where the mass donor experiences a luminosity drop just after the event (\citet{wellstein01}). Spectroscopic monitoring of star 34 would be required to test the binary scenario.

\smallskip

\begin{figure*}[t]
\centering
\includegraphics[width=0.49\textwidth]{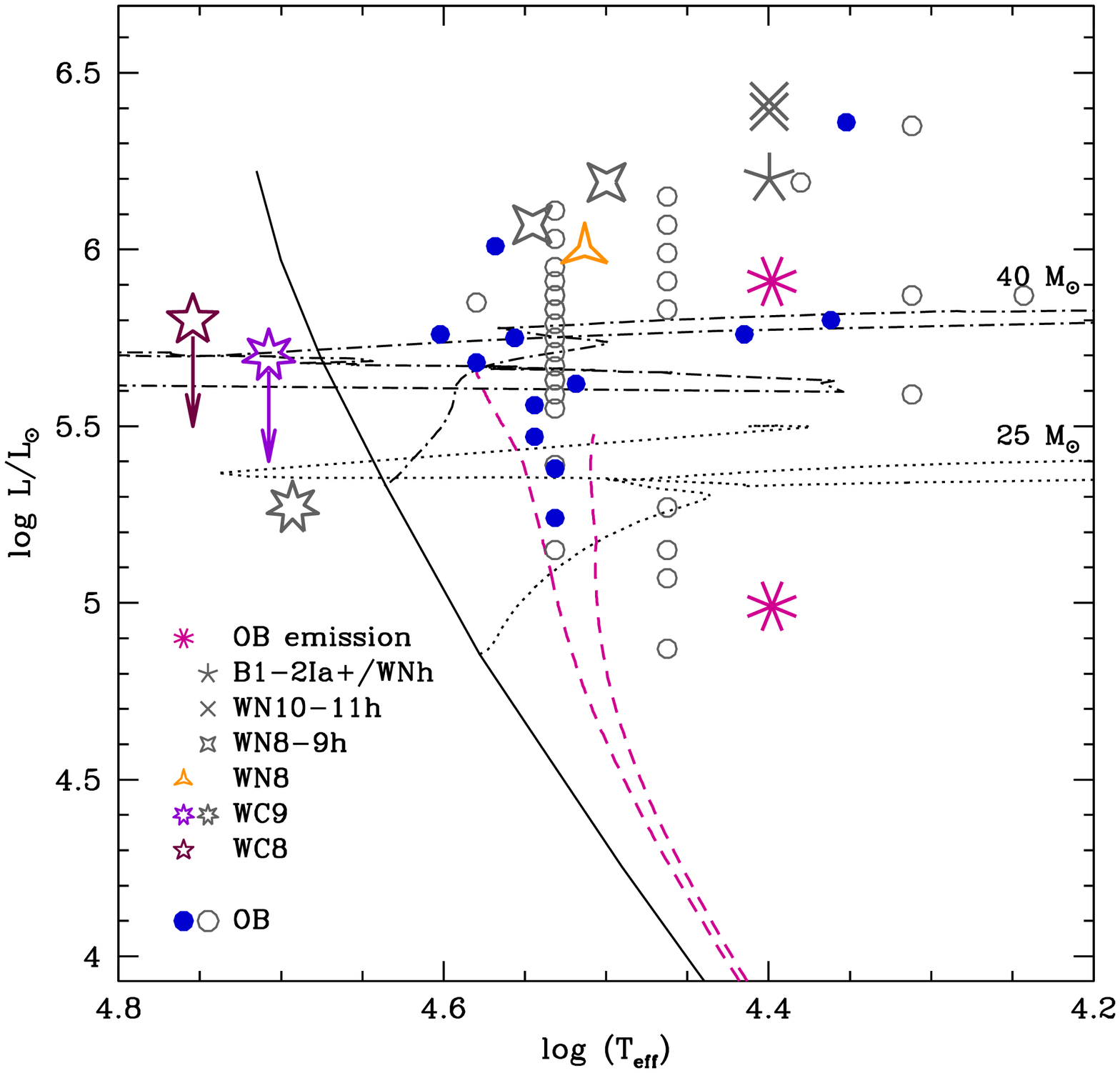}
\includegraphics[width=0.49\textwidth]{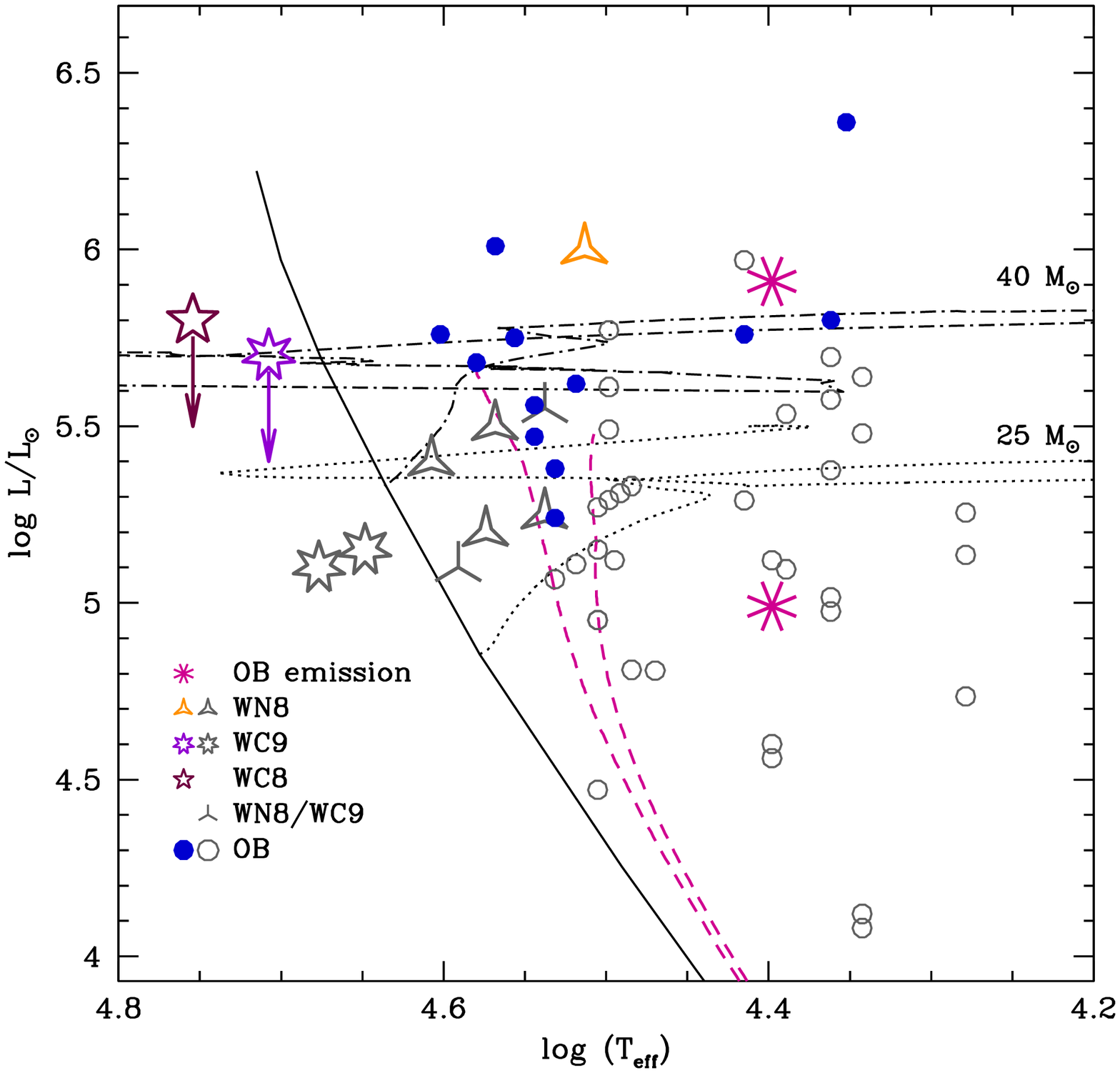}\\
\includegraphics[width=0.49\textwidth]{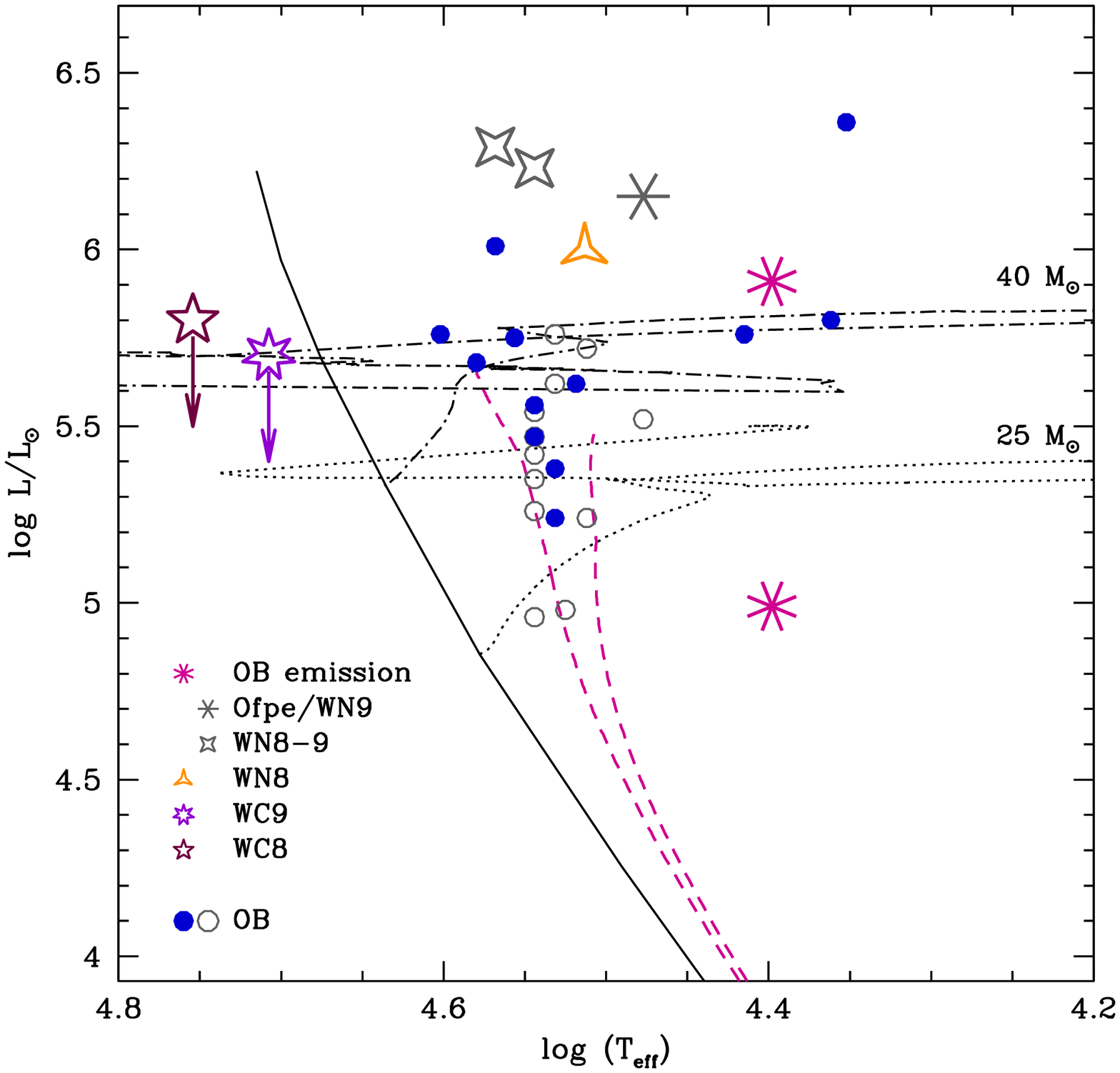}
\caption{Same as Fig.\ \ref{figHR} with OB and Wolf-Rayet stars from comparison clusters in gray: the Quintuplet \citep[top left;][]{liermann12,clark18}; the galactic center cluster \citep[top right;][]{martins07}; cluster [DBS2003]-179 \citep[bottom left;][]{borissova12}. For clarity only the 25 and 40 \msun\ tracks are plotted, as well as only the 5 and 6 Myr isochrones. Error bars have also been dropped. No distinction of luminosity class is made for OB stars.}
\label{figHRcomp}
\end{figure*}

%%------------------------------------------------------------------
\subsection{Comparison to other clusters}
\label{s_evolstat}

In Fig.\ \ref{figHRcomp} we have added the OB and WR stars of three galactic clusters: the young starburst in the central cluster of the Galaxy \citep{pgm06,martins07}; the Quintuplet cluster \citep{liermann10,liermann12}; cluster DBS2003-179 \citep{borissova12}. For the central cluster we have taken the stellar parameters from the above references. The effective temperature of the OB stars was estimated from the spectral type and the T$_{\rm eff}$-scale of \citet{msh05}. The luminosity was computed from the observed K-band magnitude, the K-band extinction \citep{pgm06}, the K-band bolometric correction \citep{mp06}, and the galactic center distance of \citet{gil09}. We have excluded objects with only upper limits on luminosities. The stellar parameters of the Quintuplet OB stars are estimated using the same method, the K-band magnitudes of \citet{liermann09}, and the revised spectral types of \citet{clark18}. The parameters of the WNh stars studied by \citet{liermann10} are also included. For these stars, we have used the spectral types of \citet{clark18} in Fig.\ \ref{figHRcomp}. The Quintuplet WC9 star GCS4 was studied by \citet{najarro17} and we adopt their average luminosity (see their Table 3) in Fig.\ \ref{figHRcomp}. Finally the stellar parameters of the stars in DBS2003-179 were taken directly from \citet{borissova12}. Table\ \ref{tab_clusters} gathers the main properties of VVV~CL074 and the comparison clusters. 

The OB stars in VVV~CL074 are located within the area covered by the OB stars of the Quintuplet. We note a rather large dispersion among the latter objects. This is nevertheless an indication that the two clusters are about the same age. Estimates for the Quintuplet are usually approximately 3-5 Myr \citep{figer99,liermann12}. This is consistent with the range 3-6 Myr we estimate for VVV~CL074. The Quintuplet cluster hosts a population of WN10-11h/B hypergiant stars, which are on average slightly cooler and more luminous than the WN8 star of VVV~CL074. The difference in temperature is qualitatively consistent with the differences between spectral types. Since both clusters have similar ages, one can speculate that the WR stars have progenitors with similar initial masses, approximately 40-60~\msun\ according to the tracks of \citet{ek12}. The WN stars in VVV~CL074 and the Quintuplet cluster are thus probably in close evolutionary states. \citet{liermann10} report a hydrogen mass fraction between 5 and 45\% for the WN10-11h star. We find such a fraction to be 13\% in the WN8 star. Hence, the WN8 phase is likely very close to the WN10-11h phase. We note that the Quintuplet  cluster also hosts a WN6 and several WC8-9 stars for which no stellar parameters have been determined \citep{figer99}, except for star GCS4, one of the five stars first discovered by \citet{nagata90} in the cluster. \citet{najarro17} performed a quantitative analysis of the infrared spectrum of this source, assuming it is a binary system made of both a WC9 and an O star. The binary nature is suggested by the pinwheel structure of the infrared emission observed by \citet{tut06}. The effective temperature obtained by \citet{najarro17} for the WC9 component is similar to what we obtain for star 13. GCS4 is less luminous than star 13 by roughly 0.5 dex, but luminosity estimates depend critically on extinction \citep[see Table 3 of][]{najarro17}, and our luminosity estimate is an upper limit. Finally, star 48 in VVV~CL074 may be a slightly lower luminosity version of the WN10-11h/B hypergiant stars of the Quintuplet cluster. They indeed share many spectroscopic properties. 

Turning to the central cluster (top right panel of Fig.\ \ref{figHRcomp}), almost all OB stars of VVV~CL074 are hotter than their galactic center counterparts. This is consistent with the older age estimated for the central cluster: 6-8 Myr according to \citet{pgm06}, compared to 3-6 Myr for VVV~CL074. The WR population of the galactic center is on average less luminous than that of VVV~CL074. If the central cluster is older, its WR population likely comes from stars with initial masses lower than the WR progenitor masses in VVV~CL074. This explains the luminosity differences. From this comparison, we conclude that WC9 and WN8 stars, which are found in both clusters, are found at different luminosities and come from stars of various initial masses.

Finally, the cluster DBS2003-179 (bottom panel of Fig.\ \ref{figHRcomp}) has a population of OB stars very similar to that of VVV~CL074 (see Table \ref{tab_clusters}). Thus, both clusters are about the same age. DBS2003-179 hosts two WN8-9 stars that have temperatures and luminosities comparable, within the error bars, to star 3 in VVV~CL074. The Ofpe/WN9 star in DBS2003-179 is also located close to star 3. The main difference between the massive stars population of the two clusters is the absence of WC stars in DBS2003-179. Otherwise, both clusters are very similar (see Table\ \ref{tab_clusters}). The presence of WN8-9 stars in both of them strengthens the case that stars with initial masses around 40-60 \msun\ go through a WN8-9 phase during their evolution.
Another similarity between VVV~CL074 and DBS2003-179 can be obtained if one re-estimates the cluster's mass using the updated stellar census of VVV~CL074. From Fig.\ \ref{figHR} we can roughly estimate the total initial mass of the stars for which we have determined the stellar parameters. One (5, 11) star(s) is located between the 20 and 25 \msun\ (respectively 25 and 40 \msun, 40 and 60 \msun) tracks. Assuming average masses of 22, 32, and 50 \msun in these mass intervals, we obtain a total mass of about 732 \msun. If we further assume that we have observed all stars more massive than 20 \msun\ (see Fig.\ \ref{figHR}) and a Kroupa IMF \citep{kroupa02} for which $\sim$10\% of the mass is in the mass range 20-120 \msun\ (see Kroupa's Table 2), we obtain a total cluster mass of about 7300 \msun, very close to the mass of  DBS2003-179.

\begin{table*}
\begin{center}
\caption{Properties of VVV~CL074 and comparison clusters. N(O) is the number of O stars, excluding stars with uncertain OB classification. N(WN) is the number of H-free and H-rich WN stars. N(WC) is the number of WC stars (including dusty ones).}
\label{tab_clusters}
\begin{tabular}{lcccccc}
\hline
Cluster          & Mass     & age     &  N(O)  &  N(WN)  &  N(WC) & References\\    
                 & [\msun]  & [Myr]   &       \\
\hline
\smallskip
VVV~CL074        & 1900--7300 & 3-6  &  10  &  1  &  2 & 1 and present study\\
Quintuplet       & $>$6300    & 3-5  &  31  &  6  & 13 & 2,3\\
Central cluster  & 15000      & 6-8  &  90  &  18 & 13 & 4,5 \\
DBS2003-179      & 7000       & 2-5  &  12  &  3  &  0 & 6,7 \\
\hline
\end{tabular}
\tablefoot{References: 1: \citet{chene13}; 2: \citet{figer99b}, 3: \citet{clark18}; 4: \citet{bartko09}; 5: \citet{pgm06}; 6: \citet{borissova08}; 7: \citet{borissova12}. }
\end{center}
\end{table*}

%%%%%%%%%%%%%%%%%%%%%%%%%%%%%%%%%%%%%%%%%%%%%%%%%%%%%%%%%%%%%%%%%%%%%%%%%%%%%%%%%%%%%%%%%%%%%%%%%%%%%%%%%%%%%%%%%%%%%%%%%%%%%%%
%%%%%%%%%%%%%%%%%%%%%%%%%%%%%%%%%%%%%%%%%%%%%%%%%%%%%%%%%%%%%%%%%%%%%%%%%%%%%%%%%%%%%%%%%%%%%%%%%%%%%%%%%%%%%%%%%%%%%%%%%%%%%%%
\section{Summary and conclusion}
\label{s_conc}

We have presented observations of the galactic cluster VVV~CL074 conducted with the integral field spectrograph SINFONI on the ESO/VLT. We have uncovered a population of 25 OB and Wolf-Rayet stars, 19 being new discoveries, based on the presence of hydrogen, helium, and sometimes carbon and nitrogen lines in their K-band spectra. Four objects classified as O or B (with a loose classification for three objects) are likely foreground stars. The remaining 15 newly discovered stars are likely cluster members. For 21 additional stars, their spectra showed strong CO absorption. These objects are most likely not physically associated to the cluster.

We estimated the stellar parameters of the WR stars and the brightest OB stars from a spectroscopic analysis with the code CMFGEN. We placed the stars in the HR diagram. The cluster hosts a population of stars formed most likely 3 to 6 Myr ago. From the relative position of the WR and OB stars we conclude that the initial mass of WN8 and WC9 stars is 40 to 60 \msun\ in VVV~CL074. We performed a comparison to three galactic massive clusters: DBS2003-179, the Quintuplet cluster, and the central cluster. VVV~CL074 has a massive star population, and thus an age, very similar to DBS2003-179, from which we infer that stars with initial masses 40-60 \msun\ go through a WN8-9 phase during their evolution. The Quintuplet cluster also has a population of a similar age and hosts WN10-11h. The comparison of the two clusters indicates a very close evolutionary state for WN10-11h and WN8 stars, which likely have an initial mass of about 40-60 \msun. Finally, the central cluster of the Galaxy is older than VVV~CL074. Its WR population comes from stars with lower initial masses compared to the progenitors of the WR stars in VVV~CL074. WN8 and WC9 stars being present in both clusters, this shows that such spectral types are encountered during the evolution of stars with a range of initial masses.

%%#####################################################################
\section*{Acknowledgments}

We acknowledge a very constructive report by the referee, Paul Crowther. We warmly thank John Hillier for making the code CMFGEN available to the community and for constant help with it. We thank Paul Crowther for discussions about the modeling of infrared spectra of WC stars, prior to submission. We gratefully acknowledge  data from the ESO Public  Survey program ID 179.B-2002  taken with  the  VISTA telescope,  and  products from  the Cambridge Astronomical Survey Unit (CASU).  
DM and JB gratefully acknowledge support provided by the BASAL  Center  for  Astrophysics and  Associated  Technologies  (CATA) through grant  PFB-06, and the  Ministry for the  Economy, Development, and Tourism, Programa Iniciativa  Cient\'ifica Milenio grant IC120009, awarded to  the Millennium Institute  of Astrophysics (MAS), and from project Fondecyt No. 1170121. SRA acknowledges the support from the FONDECYT Iniciación project Nº11171025 and the CONICYT PAI “Concurso Nacional Inserción de Capital Humano Avanzado en la Academia 2017” project PAI 79170089. ANC is supported by the Gemini Observatory, which is operated by the Association of Universities for Research in Astronomy, Inc., on behalf of the international Gemini partnership of Argentina, Brazil, Canada, Chile, and the United States of America.

%%#####################################################################
\bibliographystyle{aa}
\bibliography{vvv74}
%%#####################################################################
%%#####################################################################

\end{document}